\documentclass{aastex63}
\usepackage{amsmath}
\usepackage{wrapfig}

\newcommand{\totalvortices}{990}
\newcommand{\maskedvortices}{44}
\newcommand{\largestDeltaPobs}{$\left( 8.9 \pm 0.2 \right)\,{\rm Pa}$}
\newcommand{\largestDeltaPobssol}{65}
\newcommand{\largestGammaobssol}{20}
\newcommand{\largestGammaobs}{$\left( 99 \pm 3 \right)\,{\rm s}$}
\newcommand{\boverDactltone}{88}

\newcommand{\numICCimages}{1527}

\submitjournal{Planetary Science Journal}
\accepted{2021 Aug 30}

\shorttitle{Vortices and Meteorology from InSight}
\shortauthors{Jackson et al.}

\graphicspath{{./}{figures/}}

\begin{document}

\title{Inferring Vortex and Dust Devil Statistics from InSight}

\correspondingauthor{Brian Jackson}
\email{bjackson@boisestate.edu}

\author[0000-0002-9495-9700]{Brian Jackson}
\author{Justin Crevier}
\author{Michelle Szurgot}
\author{Ryan Battin}
\affiliation{Department of Physics\\ 
Boise State University\\ 
1910 University Drive, Boise ID 83725-1570 USA}
\author[0000-0002-7200-5682]{Cl\'{e}ment Perrin}
\affiliation{Laboratoire de Plan\'{e}tologie et G\'{e}odynamique, UMR6112, OSUNA UMS3281, Univ. Nantes, Univ. Angers, CNRS, 2 rue de la Houssini\`{e}re - BP 92208, 44322 Nantes Cedex 3, France}
\author[0000-0003-1219-0641]{S\'{e}bastien Rodriguez}
\affiliation{Universit\'{e} de Paris, Institut de Physique du Globe de Paris, CNRS, Paris, France}
\affiliation{Institut Universitaire de France, Paris, France}

\begin{abstract}
The InSight mission has operated on the surface of Mars for nearly two Earth years, returning detections of the first Marsquakes. The lander also deployed a meteorological instrument package and cameras to monitor local surface activity. These instruments have detected boundary layer phenomena, including small-scale vortices. These vortices register as short-lived, negative pressure excursions and closely resemble those that could generate dust devils. Although our analysis shows InSight encountered more than 900 vortices and collected more than 1000 images of the martian surface, no active dust devils were imaged. In spite of the lack of dust devil detections, we can leverage the vortex detections and InSight’s daily wind speed measurements to learn about the boundary layer processes that create dust devils. We discuss our analysis of InSight’s meteorological data to assess the statistics of vortex and dust devil activity. We also infer encounter distances for the vortices and, therefrom, the maximum vortex wind speeds. Surveying the available imagery, we place upper limits on what fraction of vortices carry dust (i.e., how many are bonafide dust devils) and estimate threshold wind speeds for dust lifting. Comparing our results to detections of dust devil tracks seen in space-based observations of the InSight landing site, we can also infer thresholds and frequency of track formation by vortices. Comparing vortex encounters and parameters with advective wind speeds, we find evidence that high wind speeds at InSight may have suppressed the formation of dust devils, explaining the lack of imaged dust devils.
\end{abstract}


\section{Introduction} \label{sec:Introduction}
The InSight spacecraft landed on Elysium Planitia ($4.5^\circ$ N, $135.6^\circ$ E -- \citealp{2020E&SS....701248G}) on 2018 Nov 28, carrying suites both of geophysical \citep{2020NatGe..13..183B} and meteorological instruments \citep{Banfield2019,2020NatGe..13..190B}. Since wind gusts and other atmospheric boundary layer phenomena can perturb the geophysical measurements, particularly the seismic signals, these meteorological instruments provide crucial information for mission's foci, including an exploration of Mars' interior structure and thermal state and constraining its present-day seismicity. As an added benefit, the meteorological instrumentation, the Auxiliary Payload Sensor Suite (APSS), characterize active boundary layer processes. \citet{2018SSRv..214..109S} discusses the wide range of meteorological applications and insights that the mission may yield. For instance, wind speeds measured by InSight's Temperature and Winds for InSight (TWINS) instrument may be combined with measurements of the surface (from the RADiometer instrument) and near-surface temperatures (from TWINS) to assess the accuracy and applicability of theoretical predictions of surface layer heat and momentum transport developed for Earth. InSight's cameras \citep{2018SSRv..214..105M}, the Instrument Deployment Camera (IDC) and Instrument Context Camera (ICC), may also reveal active sediment transport, observations of which, when combined with wind speeds from TWINS, can constrain threshold conditions for aeolian activity on Mars \citep{2012Natur.485..339B}.

InSight also promises to help elucidate one of the most dramatic and significant aeolian interactions on Mars, dust devils. These atmospheric apparitions arise when a convective cell draws in surrounding air which, as it conserves vorticity, spins up into a small-scale (10s to 100s of meters) whirlwind that then collects and lifts surficial dust into the atmosphere. Dust devils also frequently rearrange sediment on the martian surface, leaving long and narrow bright or dark tracks in their wakes \citep{2016SSRv..203..143R}. Observations of dust devils on Mars extend back to the Viking missions \citep{1985Sci...230..175T}, and they have since been observed in imagery from almost all landed missions to Mars. Some dust devils have even been large enough to be seen from orbiting spacecraft \citep[\emph{e.g.},][]{2011GeoRL..3824206C}.

As convective vortices, dust devils may register not just in imagery but also in meteorological time-series if a vortex passes over or near the lander -- the convective cells produce short-lived (few seconds), negative pressure excursions, accompanied by perturbations to the observed wind speed and direction \citep{2016Icar..271..326L}. Consequently, studies also going back to Viking have analyzed such meteorological data, identifying hundreds and even thousands of such encounters \citep{2003Icar..163...78R, 1999GeoRL..26.2781M, 1983JGR....8811005R, 2003JGRE..108.5133F, 2002GeoRL..29.2103M, 2016SSRv..203..277L}.

One significant drawback of such studies, though, is that, without simultaneous imagery or pyranometry, judging whether the encountered convective vortex actually carries dust is difficult. Indeed, such dustless vortices are common on both Mars and Earth, and there is no requirement that a vortex lifts dust, even when dust is available. In a terrestrial field experiment, \citet{LORENZ20151} found about 20\% of encountered vortices exhibited signatures consistent with dust loading. \citet{2016Icar..278..180S} reported 245 vortices encountered by the Mars Science Laboratory and found only two with clear signatures of dust-loading. 

Moreover, the exact conditions that allow a vortex to lift dust are unclear. Boundary layer fluid mechanics suggests inter-particle cohesion conspires with gravity to give a threshold wind velocity for dust lifting that falls with particle diameter before rising again \citep{1985wagp.book.....G}. However, seminal laboratory simulations described in \citet{2010Icar..206..306N} show instead that the smallest particles are lofted even with very small velocities. 

Worse still, exactly how much dust devils contribute to the lifting of dust into the martian atmosphere remains highly uncertain. According to \citet{2016SSRv..203...89F}, dust devils may contribute between 25\% and 75\% of the total dust flux in the martian atmosphere. Since Mars' atmosphere is so thin and provides so little greenhouse warming compared to Earth's, the aerosols suspended in Mars' atmosphere absorb and scatter significant amounts of radiation, contributing perhaps tens of degrees K of warming \citep{2002Icar..157..259S, 2004JGRE..10911006B}. Thus, accurately assessing both the amount of dust lifted by a dust devil and the frequency with which dust devils occur on Mars are critical to understanding Mars' climate and dust cycle.

In this study, we analyze the pressure and wind speed data collected by InSight's APSS both to probe the structures of individual vortices and to estimate their occurrence rates. To assess how frequently the vortices actually carry dust, we also survey the available InSight imagery. We compare our results to other recent studies of the InSight meteorological data \citep{2021Icar..35514119L, Spiga2021}. Our study differs from these previous ones in several ways: more data were made available since those studies were completed (almost 100 sols more), and we employ several novel detection and time-series analysis schemes. These differences produced some results that agree with those previous studies and some which differed. We also compared our occurrence rates to analysis of space-based observations of tracks left in the region around InSight \citep{2016Icar..266..315R, 2020GeoRL..4787234P}, which allows us to assess how frequently vortices leave tracks on the martian surface. 

We start by describing our detailed analysis of the meteorological time-series (Section \ref{sec:Meteorological Time-Series Analysis}). This section includes a description of the data themselves (Section \ref{sec:Time-Series Data}), our procedures for detecting the vortices and analyzing their pressure profiles (Section \ref{sec:Searching for Vortex Encounters and Fitting Pressure Profiles}), our procedures for analyzing the wind profiles (Section \ref{sec:Fitting Wind Profiles}), and the resulting vortex statistics (Section \ref{sec:Vortex Statistics}). Finally, we discuss our estimates for the intrinsic vortex occurrence rates (Section \ref{sec:Inferring Areal Occurrence Rate from the Time-Series Analysis}). In the next section, we discuss our analysis of the ICC imagery to infer the intrinsic occurrence rates for dust devils themselves (Section \ref{sec:Image Analysis}). A detailed comparison of our results to previous studies comes next (Section \ref{sec:Discussion}), followed by a discussion of caveats and future work (Section \ref{sec:Conclusions}). Two appendices follow that provide details on the statistics of our vortex detection scheme and on our model for determining the geometry and uncertainties for each encounter between the InSight lander and a vortex from the observed vortex parameters. 

\section{Meteorological Time-Series Analysis}
\label{sec:Meteorological Time-Series Analysis}
In this section, we describe our analysis of the pressure and wind speed time-series. We first describe the data themselves. Next, we discuss our search for vortex signals using the pressure time-series and explore its biases and completeness, which \citet{2015JGRE..120..401J} showed are important for inferring the underlying occurrence rates from the observations. Next, we describe how we modeled the pressure and wind speed profiles of individual vortex encounters, which, as we show, allows us to determine not just the observed pressure and wind excursions but also to infer the encounter geometries and intrinsic vortex parameters \citep[\emph{cf.}][]{2016Icar..271..326L}. We also discuss the statistical distributions and correlations of vortex parameters. 

\subsection{Time-Series Data}
\label{sec:Time-Series Data}
The pressure measurements from the APSS are taken at $10\, {\rm Hz}$ with a nominal precision of $50\,{\rm mPa\ Hz^{-1/2}}$ or better, much higher frequency and precision than available from some previous Mars landers \citep[\emph{e.g.},][]{2010JGRE..115.0E16E}. As we discuss below, turbulent excursions give an effective scatter in the pressure data between $0.2$ and $0.5\,{\rm Pa}$, depending on ambient conditions. In any case, such specs make APSS ideal for studying turbulent signals in the martian boundary layer \citep{2018SSRv..214..109S}. APSS has measured pressures nearly non-stop since sol 14 of the mission, and for our study, we considered data up through sol 477 of the mission, amounting to almost 82 GB. The data are available from NASA's Atmosphere's PDS Node. PDS provides several sets of data files for APSS, and we used the CSV files in the ``data\_calibrated'' folder. These data files are different from the raw data files because they include a temperature-dependent calibration -- see https://atmos.nmsu.edu/PDS/data/PDS4/InSight/ps\_bundle/document/pressure\_processing.pdf for details.

Wind data come from the TWINS instrument, the sensors for which sit on booms located on the InSight platform and facing opposite directions about a meter above the solar panels \citep{2020NatGe..13..190B}. TWINS acquires data at $0.1\, {\rm Hz}$ and $1\, {\rm Hz}$ with an accuracy of $1\, {\rm m\ s^{-1}}$ for wind speed and $22.5^\circ$ for wind direction. In fitting the vortex wind profiles, we do not consider the TWINS wind direction data. Although these data could help us to reconstruct the encounter geometries and determine cyclonicity (clockwise or counter-clockwise rotation), the precision of $22.5^\circ$ is insufficient to provide robust constraints (though the directional data are useful for studying other boundary layer processes). Moreover, our analysis requires only the magnitude of the vortex wind speeds, not the direction. TWINS has also provided a dataset spanning nearly the whole time from sol 14 to 477, amounting to more than 18 GB. For our work here, we focus on the higher time resolution ($1\, {\rm Hz}$) wind data, which are somewhat more limited in extent, often only spanning the mid- to late afternoon for a given sol. The wind measurements involve modeled reconstructions, as described in \citet{Banfield2019}, and we used the CSV files in the ``data\_derived'' folder. A higher resolution dataset ($20\,{\rm Hz}$) is labeled as ``modelevent'' on PDS, but it is more limited in extent. So we opted to use the lower-resolution data. ``Derived'' data involve modeling out instrumental effects to achieve a (presumably) more accurate representation of the wind field; ``Calibrated'' data involve converting the raw instrument measurements to physical quantities. See https://atmos.nmsu.edu/PDS/data/PDS4/InSight/twins\_bundle/document/twinsps\_dp\_sis\_issue10.pdf for more details.

Vortex encounters can also produce excursions in ambient temperature as the warm core passes over the sensor \citep{2016SSRv..203...39M}, and APSS does return air temperature data. However, we do not model these time-series since the temperature data show small or negligible excursions during the encounters \citep{2021Icar..35514119L}. In any case, the temperatures would be expected to simply mirror the pressure excursions \citep{2016Icar..271..326L}.

\subsection{Searching for Vortex Encounters and Fitting Pressure Profiles}
\label{sec:Searching for Vortex Encounters and Fitting Pressure Profiles}
\begin{figure}
    \centering
    \includegraphics[width=\textwidth]{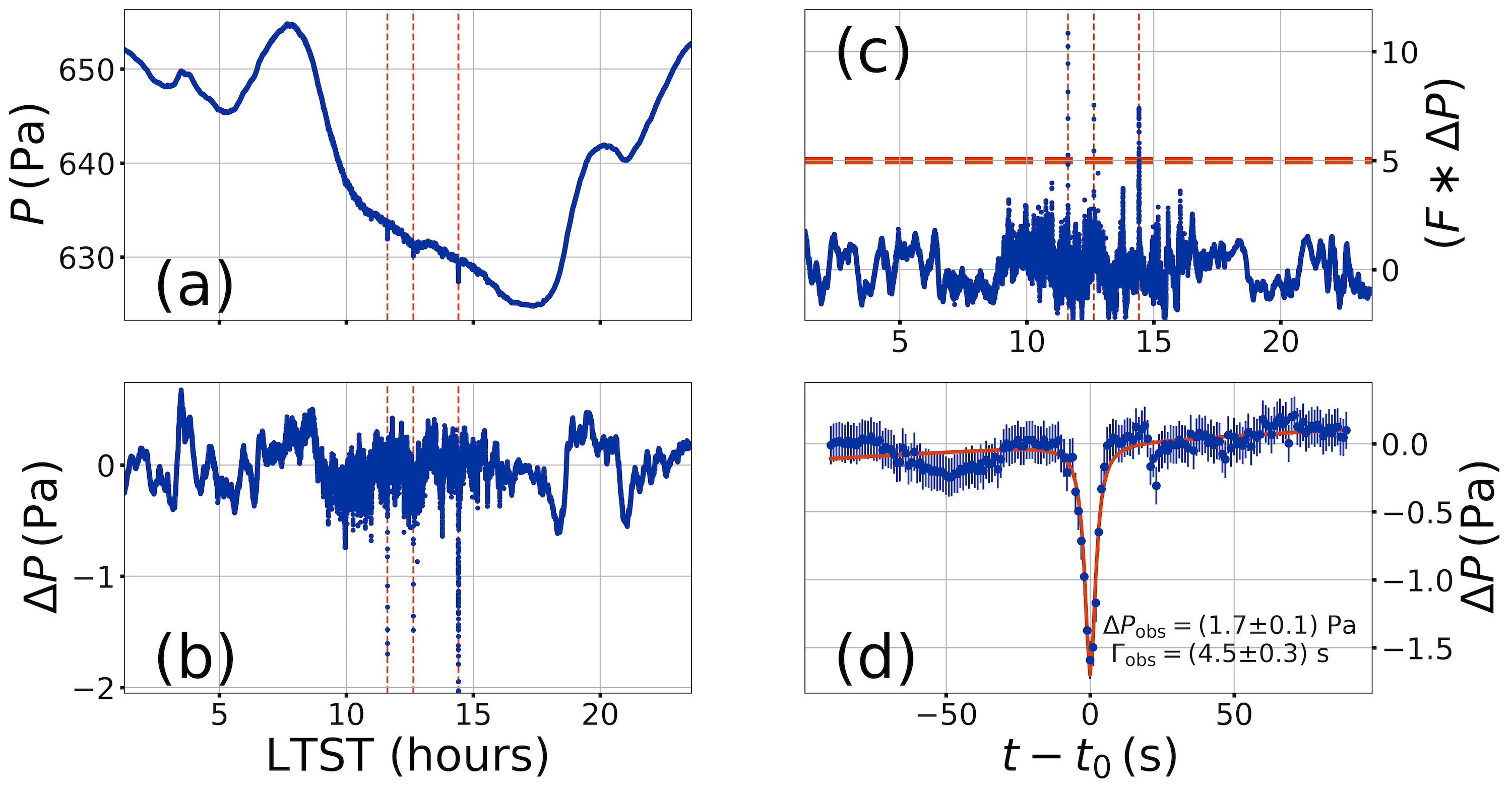}
    \caption{(a) The pressure time-series for sol 395, as blue dots. The vertical orange lines highlight the detected vortex signals. (b) The time-series after application of the mean boxcar filter. Apparent by eye, the scatter in the time-series increases around mid-day. (c) Convolution of the matched filter with the time-series in (b). The horizontal dashed orange line shows the detection threshold of 5. (d) A model fit (solid orange line) to the deepest vortex discovered on sol 395, along with the model fit parameters -- each point's uncertainty is calculated via $1.4826\ \times$ the median absolute deviation in the window centered on that point.}
    \label{fig:data_conditioning_and_fit}
\end{figure}

Both the pressure and wind speed data exhibit turbulent excursions that constitute a source of non-Gaussian noise and complicate our search for vortex encounters. However, the pressure data are both more plentiful and less affected by these excursions, so we search for encounters using the pressure time-series. This approach resembles many prior studies, including previous analyses of InSight data \citep{Spiga2021, 2021Icar..35514119L}. Figure \ref{fig:data_conditioning_and_fit} depicts our search process in graphical form, and panel (a) shows the raw pressure time-series for sol 395, a representative sol. The vertical dashed orange lines show the vortex signals, whose detection we describe next. Any time-series analysis scheme will unavoidably involve selection effects that can skew the recovered population of signals \citep{2018Icar..299..166J}, and we explore biases of our detection scheme and how those influence the final recovered population of vortices in the Appendix \ref{sec:Vortex Recovery Statistics}. 

To suppress longer-term signals and facilitate detection of the vortices, we apply a mean boxcar filter with a window size $W$ before sifting the data for vortices. Figure \ref{fig:data_conditioning_and_fit}(b) shows the resulting detrended time-series $\Delta P$. Based on the analysis described in the Appendix \ref{sec:Vortex Recovery Statistics}, we chose $W = 3000\,{\rm s}$. 

Next, we employ a matched filter approach \citep[][ch.~13]{Press2007} using a normalized Lorentzian profile with a known FWHW, $\Gamma$; that is, we march a Lorentzian profile, point-by-point, across the time series, convolving it with the time-series. Based on our analysis (Appendix \ref{sec:Vortex Recovery Statistics}), we chose $\Gamma = 1\,{\rm s}$. This process produces the equivalent of a spectrum, with large positive spikes when the filter encounters other Lorentzian-like signals. We subtract the median value from this raw spectrum and then normalize it by the standard deviation (as estimated by $1.4826\ \times$ the median absolute deviation -- \citealp{doi:10.1080/01621459.1993.10476408}). We consider peaks rising above the detection threshold of 5 to be possible vortices. Figure \ref{fig:data_conditioning_and_fit}(c) shows this normalized spectrum for the time-series in panel (b), along with the peaks raising above $F \ast P = 5$ (vertical dashed orange lines). 

Finally, considering the original, undetrended time-series (\emph{e.g.}, Figure \ref{fig:data_conditioning_and_fit}a), we use the Levenberg-Marquardt algorithm \citep[\emph{cf.}][]{Press2007} to fit the time-series in a window 30-FWHMs wide around each vortex signal. As in previous work \citep[\emph{e.g.},][]{2016JGRE..121.1514K}, we assume the pressure structures of vortices are accurately represented by a steady-state modified Lorentzian profile,

\begin{equation}
    \Delta P(r) = -\frac{\Delta P_{\rm act}}{1 + \left( \frac{r}{D_{\rm act}/2} \right)^2}\label{eqn:radial_lorentzian_profile}
\end{equation}
where $r$ the radial distance of the InSight sensors from the vortex center, $P_{\rm act}$ the pressure excursion at the vortex center, and $D_{\rm act}$ is taken as the vortex diameter. As a function of time $t$, $r(t)$ is given by
\begin{equation}
    r(t) = \sqrt{b^2 + U^2 \left( t - t_0 \right)^2}\label{eqn:radial_distance}
\end{equation}
where $t_0$ is the time of closest approach. This scheme assumes the vortex travels past the sensor on a linear trajectory with a unidirectional and constant velocity $U$ (at least during the course of a single encounter). The closest approach distance $b$ between the vortex center and the InSight sensors is usually greater than zero. As a result, the minimum observed in the pressure time-series $\Delta P_{\rm obs}$ is not usually as deep as the actual pressure excursion at the vortex center \citep{2018Icar..299..166J, 2019Icar..317..209K}. The pressure time-series for a vortex encounter can therefore be represented by 
\begin{equation}
    \Delta P(t) = \frac{-\Delta P_{\rm obs}}{1 + \left( \frac{ t - t_0 }{\Gamma_{\rm obs}/2} \right)^2}\label{eqn:Lorentzian_profile}
\end{equation}
where $\Gamma_{\rm obs}$ is the observed profile full-width/half-max (FWHM). With these definitions, $\Gamma_{\rm obs} = U^{-1} \sqrt{D_{\rm act}^2 + \left( 2 b \right)^2}$. 

We fit this profile combined with a linear trend, returning best-fit parameters $t_0$, $\Delta P_{\rm obs}$, and $\Gamma_{\rm obs}$, as well as the background slope and intercept. Fitting such a model to the original, undetrended data instead of the detrended time-series avoids the distorting effect of the boxcar filter on the vortex signal while taking into account any background trend. Figure \ref{fig:data_conditioning_and_fit}(d) shows such a fit.

\subsection{Fitting Wind Profiles}
\label{sec:Fitting Wind Profiles}
To assess the intrinsic vortex properties and encounter geometries, we also fit wind speed profiles to the wind speed time-series that coincide in time with the peaks found by our search. The wind speed signal consists both of an ambient wind $U$ and the vortex wind $V(r)$, which is a function of radial distance (and therefore time). For the Vatistas vortex model \citep{1991ExFl...11...73V}, $V(r)$ is given by
\begin{equation}
    V(r) = V_{\rm act} \frac{2 \left( \frac{r}{D_{\rm act}/2} \right) }{1 + \left( \frac{r}{D_{\rm act}/2} \right)^2}
\end{equation}
where $V_{\rm act}$ is the tangential wind speed at the vortex diameter. Similar to the pressure signal, a non-zero $b$ means the maximum wind speed encountered $V_{\rm obs}$ is less than the actual maximum at the vortex diameter. We model the observed vortex wind profile as
\begin{equation}
    V(t) = V_{\rm obs} \frac{\sqrt{1 + \left( U_1/b \right)^2 \left( t - t_0 \right)^2}}{1 + \left(\frac{t - t_0}{\Gamma_{\rm obs}/2}\right)^2}\label{eqn:wind_profile}
\end{equation}
where $U_1$ is the ambient wind speed before the encounter and which we take as the advective speed for the vortex. Strictly, this model breaks down for $b = 0$, but such an encounter is statistically unlikely \citep{2018Icar..299..166J}. Many of the wind signals exhibit a different ambient wind speed before the encounter ($U_1$) than after the encounter ($U_2$). Thus, the total wind speed observed $W(t)$ involves the vector sum of the ambient wind and vortex wind and is given by 
\begin{equation}
    W(t) = 
    \begin{array}{ll}
            \sqrt{V^2 + 2 U_1 V \cos \theta + U_1^2}, & \left( t - t_0 \right) \leq 0\\
            \sqrt{V^2 + 2 U_2 V \cos \theta + U_2^2}, & \left( t - t_0 \right) > 0\\
    \end{array} \label{eqn:total_wind_speed}
\end{equation}
where $\cos \theta = b/\sqrt{b^2 + \left( U_{1/2} \right)^2\left( t - t_0 \right)^2}$. 

We fit the pressure and wind speed profiles for each encounter in two separate steps -- first, the pressure, then the wind speed. In so doing, we hold the $\Gamma_{\rm obs}$- and $t_0$-values fixed from the pressure profile fit. To fit the wind profiles, we estimate $U_1$ and $U_2$ by finding the median wind speed $W(t)$ between $3$ and $5\times\Gamma_{\rm obs}$ before and after the encounter and then hold these values fixed as we fit $V$. Experimentation showed this approach most frequently gave reasonable results, and Figure \ref{fig:vortices_and_windspeed} shows several examples of the profile fits. 

As we show in Appendix \ref{sec:Inferring Encounter Geometries from the Pressure and Velocity Profiles}, fitting both $\Delta P_{\rm obs}$ and $V_{\rm obs}$ and assuming a balance between centrifugal and pressure gradient accelerations (i.e., cyclostrophic balance) allows us to estimate $\Delta P_{\rm act}$ and $V_{\rm act}$, along with the encounter distance $b$. To check this approach, we applied these models to many synthetic vortex encounters for a range of encounter geometries, vortex parameters, and time-series noise representative of the observed values. We found that, for encounters with $b \lesssim D_{\rm act}$, we were able to recover the assumed parameters to within 50\% for the majority of cases. For encounters farther than that, the signals were often lost in the noise. We also required that the estimated $V_{\rm obs}$ exceed the scatter in the wind speed data $\sigma$ by a factor of three to ensure a robust detection. Future work should explore more robust approaches. In what follows, we initially retain all vortices detected, regardless of their best-fit $b$-values, since the best-fit $\Delta P_{\rm obs}$- and $\Gamma_{\rm obs}$ are independent of $b$, but for results later on that depend on $b$, we dropped those vortices with $b > D_{\rm act}$ and $V_{\rm obs}/\sigma < 3$, leaving \boverDactltone\ vortices. That transition is clearly indicated in the narrative below.

\begin{figure}
\centering
\includegraphics[width=0.48\textwidth]{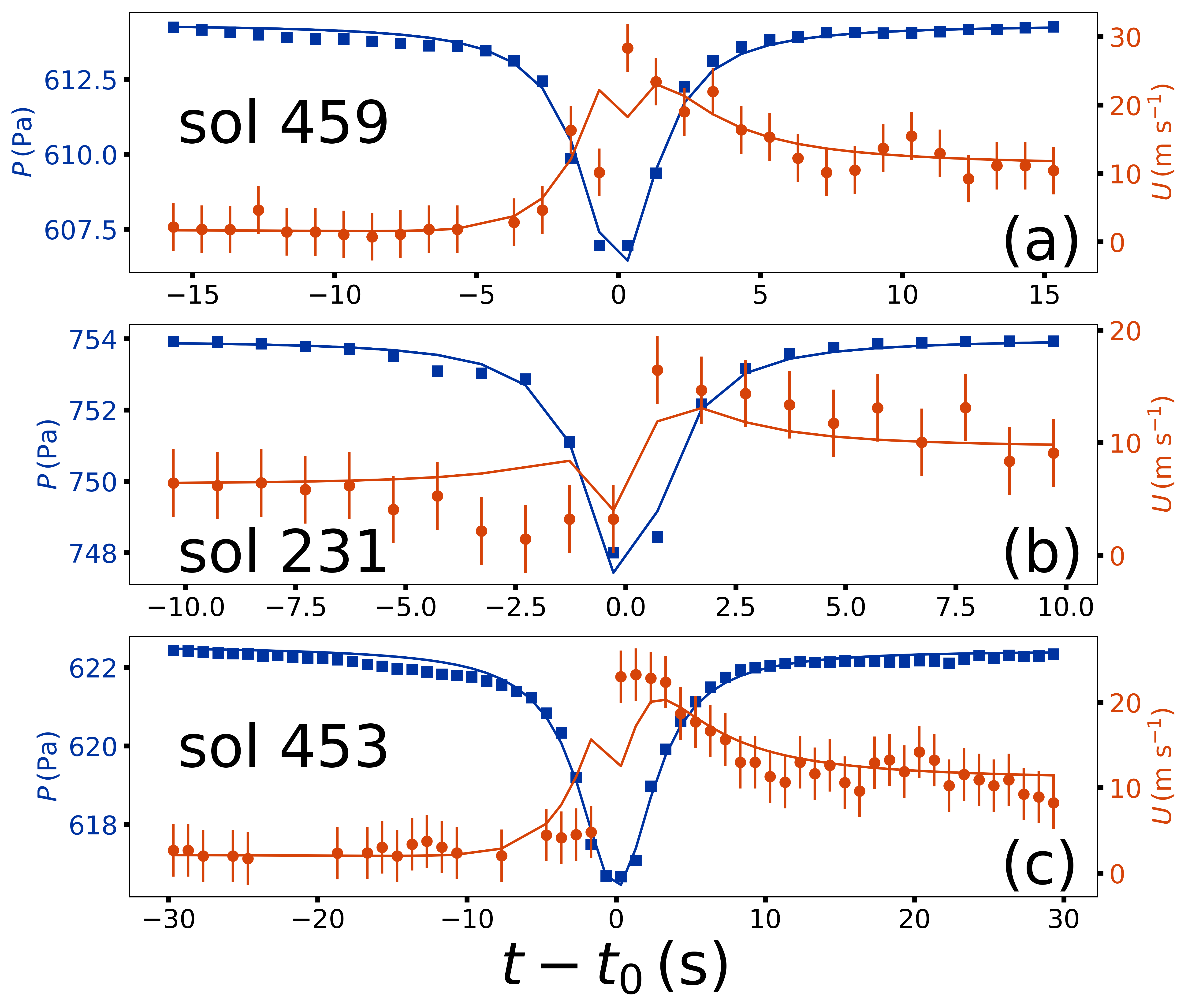}
\caption{Pressure data (blue dots) and models (blue lines) and horizontal windspeed data (orange dots) and models (orange lines). \label{fig:vortices_and_windspeed}}
\end{figure}

\subsection{Vortex Statistics}
\label{sec:Vortex Statistics}
\begin{figure}
    \centering
    \includegraphics[width=\textwidth]{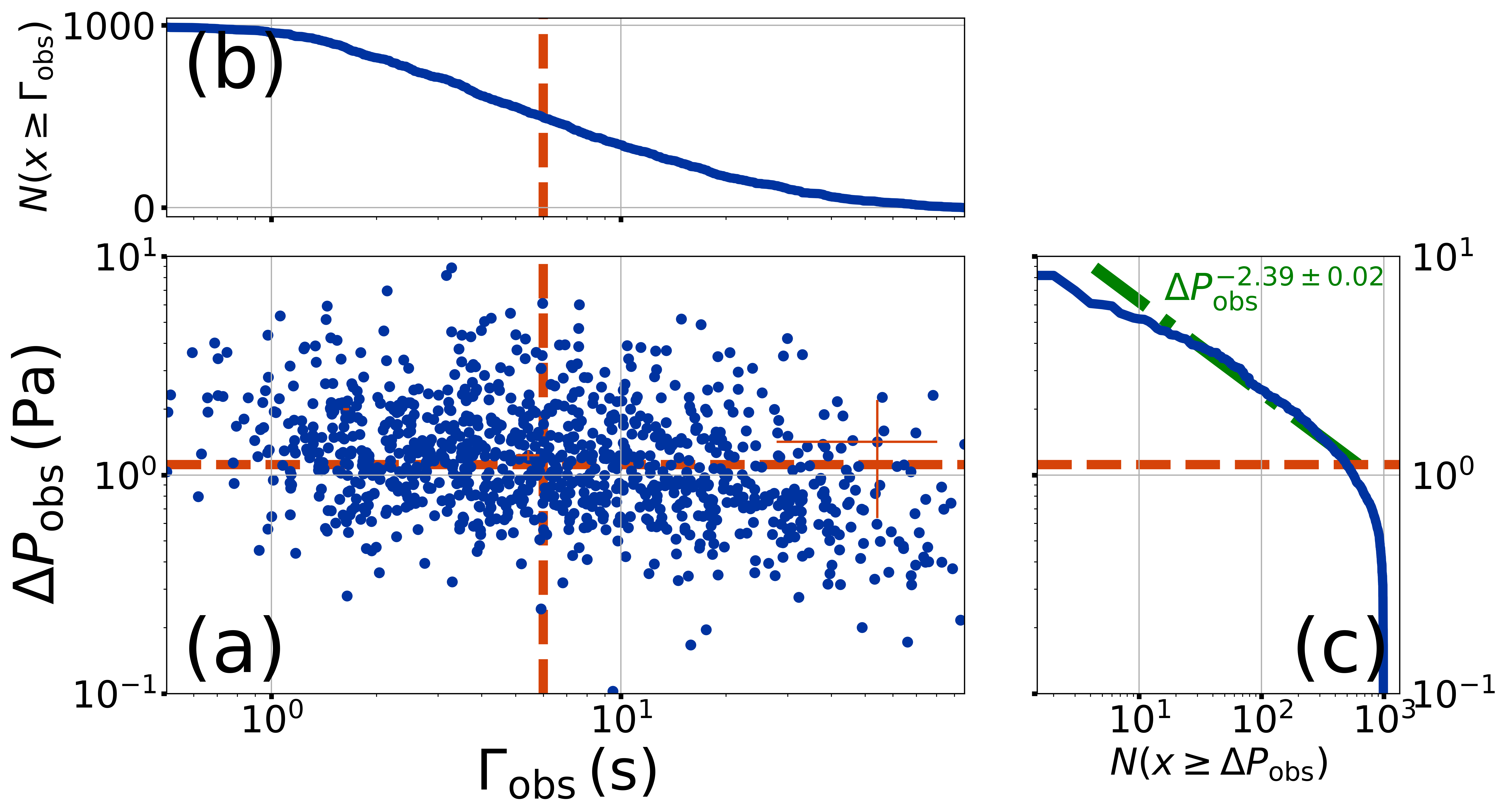}
    \caption{(a) The best-fit $\Delta P_{\rm obs}$- and $\Gamma_{\rm obs}$-values (blue dots). Representative error bars are shown as orange crosses. (b) Cumulative histogram of $\Gamma_{\rm obs}$-values, along with the median value ($\Gamma_{\rm obs} = \left( 9.3 \pm 0.2 \right)\,{\rm s}$) shown by the dashed, orange line. (c) Cumulative histogram of $\Delta P_{\rm obs}$-values, along with the median value ($\Delta P_{\rm obs} = \left( 1.13 \pm 0.03 \right)\,{\rm Pa}$) shown by the dashed, orange line. The dash-dotted green line shows a power-law fit to the histogram, with $N \propto \Delta P_{\rm obs}^{-2.39\pm0.02}$.}
    \label{fig:DeltaPobs_vs_Gammaobs}
\end{figure}

Figure \ref{fig:DeltaPobs_vs_Gammaobs}(a) shows our collection of best-fit $\Gamma_{\rm obs}$- and $\Delta P_{\rm obs}$-values, along with their respective cumulative histograms. Inspecting an initial tranche of detections by-hand, we found that vortices with best-fit $\Gamma_{\rm obs} > 100\,{\rm s}$ and/or $\Delta P_{\rm obs} < 0.1\,{\rm Pa}$ tended not to resemble true vortices but instead appeared simply to be incoherent pressure excursions, so we dropped \maskedvortices\ of these initial detections, leaving the \totalvortices\ vortex signals depicted in Figure \ref{fig:DeltaPobs_vs_Gammaobs}. The largest $\Delta P_{\rm obs}$-value we found was \largestDeltaPobs\ on sol \largestDeltaPobssol, which seems to correspond to the deepest vortex reported in \citet{Spiga2021}. The longest-duration vortex occurred on sol \largestGammaobssol\ and lasted \largestGammaobs. The median values are $\Gamma_{\rm obs} = \left( 9.3 \pm 0.2 \right)\,{\rm s}$ and $\Delta P_{\rm obs} = \left( 1.13 \pm 0.03 \right)\,{\rm Pa}$ (indicated by the dashed, orange lines in Figure \ref{fig:DeltaPobs_vs_Gammaobs}). As evident in previous analyses of Mars lander pressure time-series \citep[\emph{e.g.},][]{2010JGRE..115.0E16E}, there is a marked absence of long-duration/deep (i.e., large $\Delta P_{\rm obs}$) vortices. This absence simply reflects the miss distance effect: most encounters between the barometer and vortex occur some distance from the vortex center ($b > 0$ as in Equation \ref{eqn:radial_distance}), where the pressure profile is more shallow and of longer duration \citep{2018Icar..299..166J}.

The flattening of the cumulative histogram for $\Delta P_{\rm obs}$ (Figure \ref{fig:DeltaPobs_vs_Gammaobs}c) near $1.1\,{\rm Pa}$ indicates a decline in the number of detected vortices below that value. This decline occurs, at least in part, because of difficulty detecting these more shallow signals against noise \citep{2018Icar..299..166J}. Given the possible strong dependence of dust-lifting on $\Delta P$, the exact form of the histogram of $\Delta P_{\rm obs}$-values is critical for evaluating the population's atmospheric influence. We fit a power-law to the cumulative histogram with an exponent $\gamma = -2.39\pm0.02$, which indicates the differential histogram has an exponent $\approx -3.39$. This exponent is reasonably consistent with the values reported in \citet{Spiga2021} and \citet{2021Icar..35514119L} and with the value reported in \citet{2018Icar..299..166J} for the vortices detected by the Phoenix Mission. (On a related note, since the best-fit exponent for a differential histogram can depend on the chosen binning, we suggest such analyses use cumulative histograms or a data-informed scheme for binning -- \citealp{2016SSRv..203..277L}.)

\begin{figure}
    \centering
    \includegraphics[width=\textwidth]{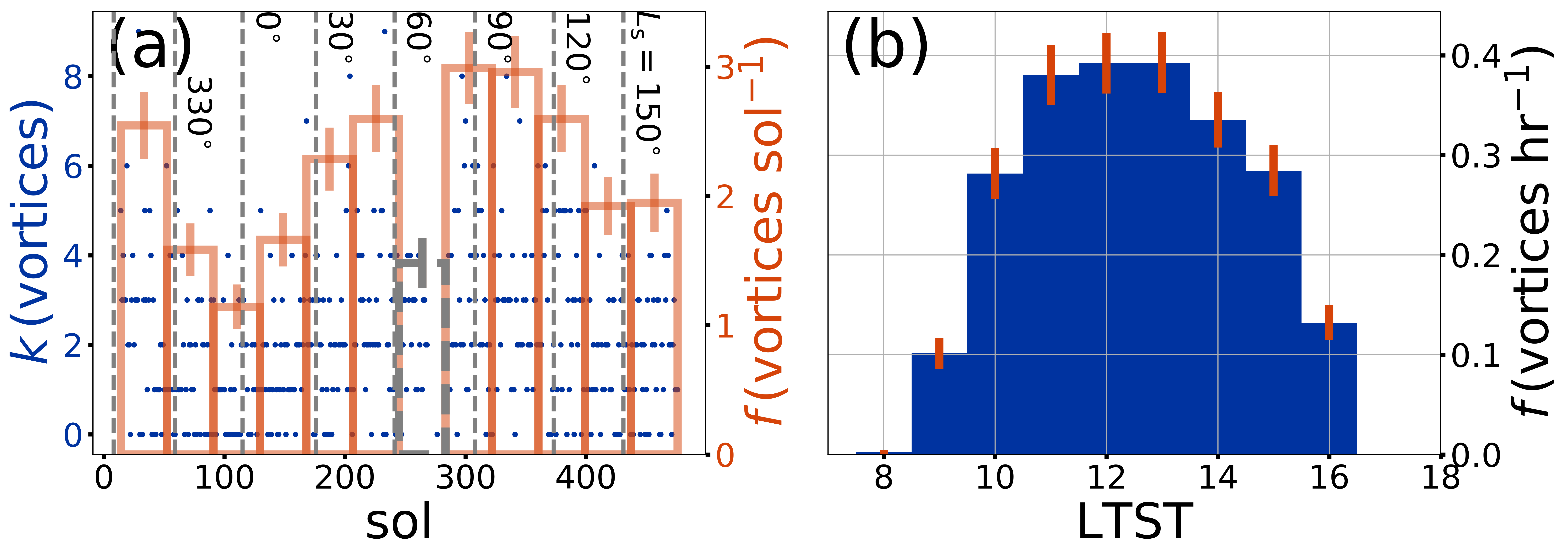}
    \caption{(a) The number of vortices during each sol $N$ (blue dots) along with the values for $L_{\rm s}$ (dashed, grey lines). (Northern spring starts at $L_{\rm s} = 90^\circ$.) The orange histogram shows the number of vortices per sol for bins 38.5 sols wide. The grey, dashed box shows the bin affected by solar conjunction when data were not available between sols 270 and 283. (b) The number of vortices per hour of local true solar time (LTST). Orange lines in both panels show Poisson error bars.}
    \label{fig:sol_and_t0_histograms}
\end{figure}

The vortices also exhibit sol-to-sol and hour-to-hour variation, as illustrated in Figure \ref{fig:sol_and_t0_histograms}. Panel (a) shows that the maximum daily number of vortices occurred on sol 204 of the mission, in the middle of northern spring, while the next maximum occurs on sol 300, near the beginning of northern summer. The dip around sol 270 (shown with a grey, dashed box) is artificial because it corresponds to solar conjunction when data are not available. Although \citet{Spiga2021} used a different procedure, that study found a broadly similar pattern -- a dip in the occurrence rate centered on sol 100, with an increase going into sol 200 and beyond. Regarding the hour-to-hour variations, panel (b) shows that the maximum occurs between 12:00 and 13:00 LTST, with a nearly symmetric decline to either side. For this calculation, we totaled up the number of hours (and fractions thereof) during which pressure time-series were reported and then divided the number of vortices encountered during each of those hours over all sols by that total. These results contrast somewhat with those of \citet{Spiga2021}, which found a peak in vortex occurrence between 11:00 and 12:00 LTST. 

\begin{figure}
    \centering
    \includegraphics[width=\textwidth]{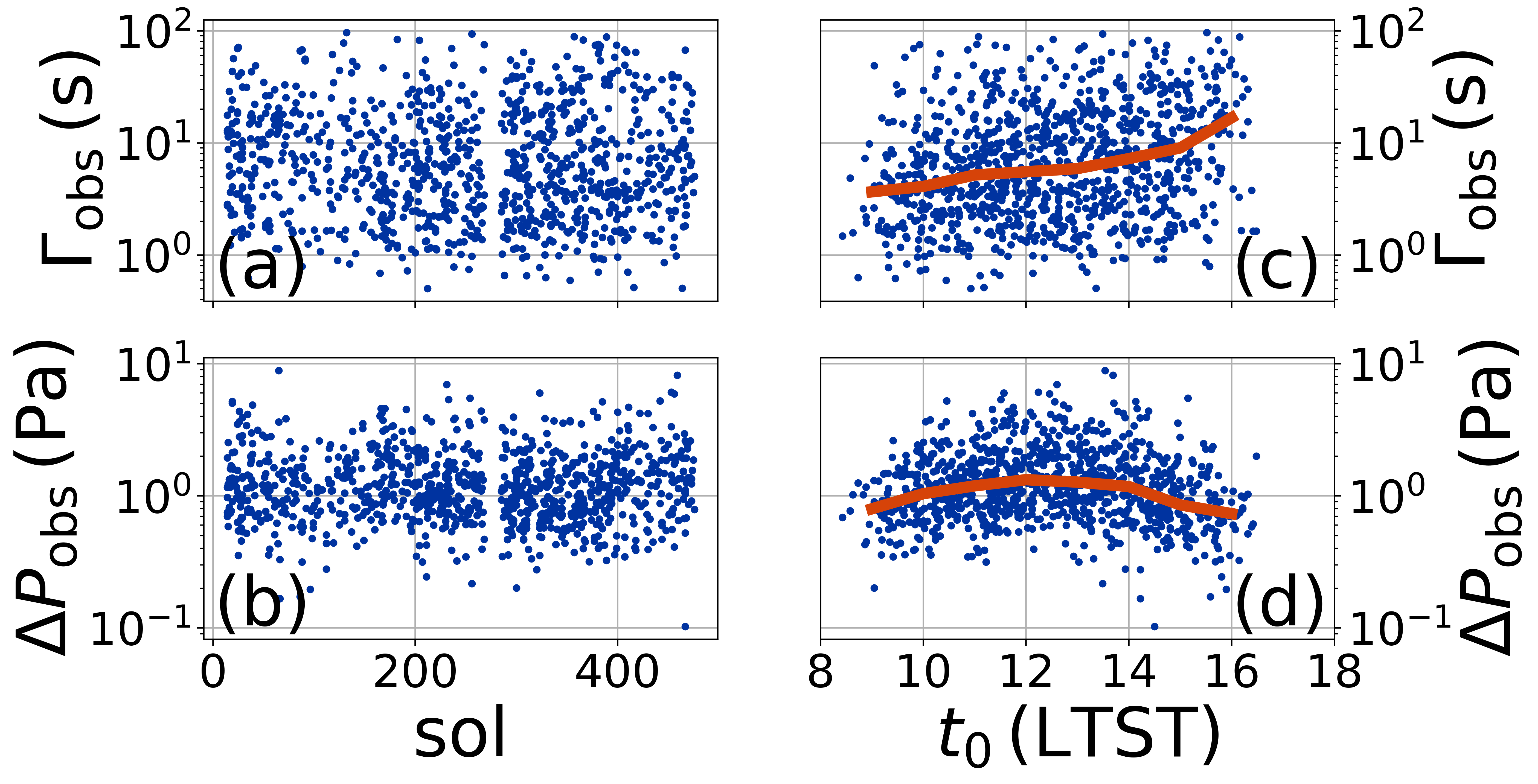}
    \caption{(a)/(b) Distributions of $\Gamma_{\rm obs}$ and $\Delta P_{\rm obs}$ by sol. (c)/(d) Distributions of $\Gamma_{\rm obs}$ and $\Delta P_{\rm obs}$ by time-of-day $t_0$. The orange lines show the median values from binning by hour.}
    \label{fig:Gammaobs_DeltaPobs_vs_TOD_and_sol}
\end{figure}

The $\Gamma_{\rm obs}$- and $\Delta P_{\rm obs}$-values also exhibit trends as well. Figure \ref{fig:Gammaobs_DeltaPobs_vs_TOD_and_sol}(c) shows that, binned by the hour, the median value of $\Gamma_{\rm obs}$ steadily increases from early morning to late afternoon. Figure \ref{fig:Gammaobs_DeltaPobs_vs_TOD_and_sol}(d) shows that median value for $\Delta P_{\rm obs}$ peaks around 12:00 LTST at $\left( 1.3\pm0.05 \right)\,{\rm Pa}$ with minimum values at either end of about $\left( 0.7\pm 0.1 \right)\,{\rm Pa}$ (where the uncertainties come from the error of the median in the corresponding bin). However, the distributions for both values also involve considerable scatter as well. The values binned by sol (Figures \ref{fig:Gammaobs_DeltaPobs_vs_TOD_and_sol}a and b) show no obvious trends. 

Presumably, these putative hour-to-hour trends reflect the influence of variable ambient conditions, but these data alone are not sufficient to judge what influence: an increase in $\Gamma_{\rm obs}$ could result either from intrinsically larger vortices late in the day or lower wind speeds $U$ advecting the same size vortices past the sensor. Fortunately, the TWINS wind speed data can shed light on this issue. We estimated the advection speed by taking the median value $U$ between 5- and 3-$\Gamma_{\rm obs}$ before the encounter time $t_0$. This approach returns a wind speed close enough in time to be a plausible estimate of the advection speed but early enough that the vortex did not significantly influence the measurement. 

\begin{figure}
    \centering
    \includegraphics[width=\textwidth]{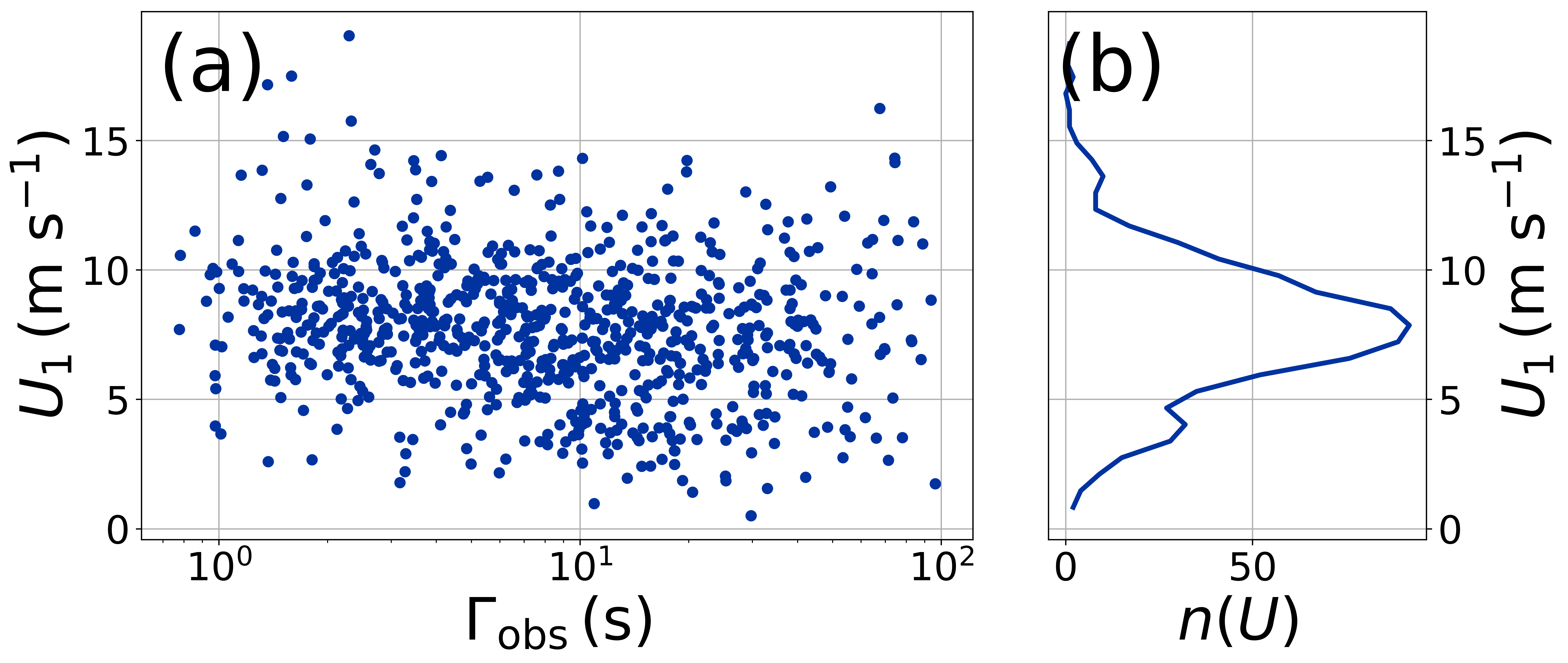}
    \caption{(a) Distribution of vortex-associated wind speeds $U_1$ as a function of the vortex duration $\Gamma_{\rm obs}$. Panel (b) shows a differential (as opposed to cumulative) histogram of $U_1$-values.}
    \label{fig:U1_vs_Gamma_hist}
\end{figure}

Figure \ref{fig:U1_vs_Gamma_hist} shows the distribution of advective wind speeds associated with the observed vortices; the maximum, median, and minimum values are $19.1\pm2.3$, $7.6\pm1.0$, and $0.5\pm0.2 \,{\rm m\ s^{-1}}$, respectively. Panel (a) suggests an anti-correlation between $U_1$ and $\Gamma_{\rm obs}$, as we might expect if trends in $\Gamma_{\rm obs}$ reflect a change in advection speed rather than simply variations in vortex diameter. A Pearson-r test confirms an anti-correlation, albeit a weak one with $r = -0.07$, at a significance $p < 0.05$. Of course, the fact that $\Gamma_{\rm obs}$ spans two orders of magnitude, while $U_1$ spans only a factor of about three indicates that variations in vortex diameter play an important role. 

Next, we couple our pressure profile and wind profile fits (Equations \ref{eqn:Lorentzian_profile} and \ref{eqn:wind_profile}) to infer the actual central pressure excursions $\Delta P_{\rm act}$ and vortex diameters $D_{\rm act}$, as well as the eyewall wind velocities $V_{\rm act}$. The wind speed data showed considerable turbulent excursions not captured by our model, so as we fit the wind speed models, we inflated the data point and model parameter uncertainties by multiplying by the square root of the reduced $\chi^2$-values, effectively imposing $\chi^2 = 1$ \citep[\emph{cf.}][]{Press2007}. We also propagated the uncertainties from the pressure and wind profile fit parameters into uncertainties on the actual parameters, as described in Appendix \ref{sec:Inferring Encounter Geometries from the Pressure and Velocity Profiles}. For these results, we only included encounters for which the inferred $b/D_{\rm act} \leq 1$, $V_{\rm obs}/\sigma \geq 3$, and with well-defined uncertainties on the actual values -- we found \boverDactltone\ such vortex encounters. As described in Section \ref{sec:Fitting Wind Profiles} above, numerical experimentation with synthetic datasets bolstered this approach.

\begin{figure}
    \centering
    \includegraphics[width=\textwidth]{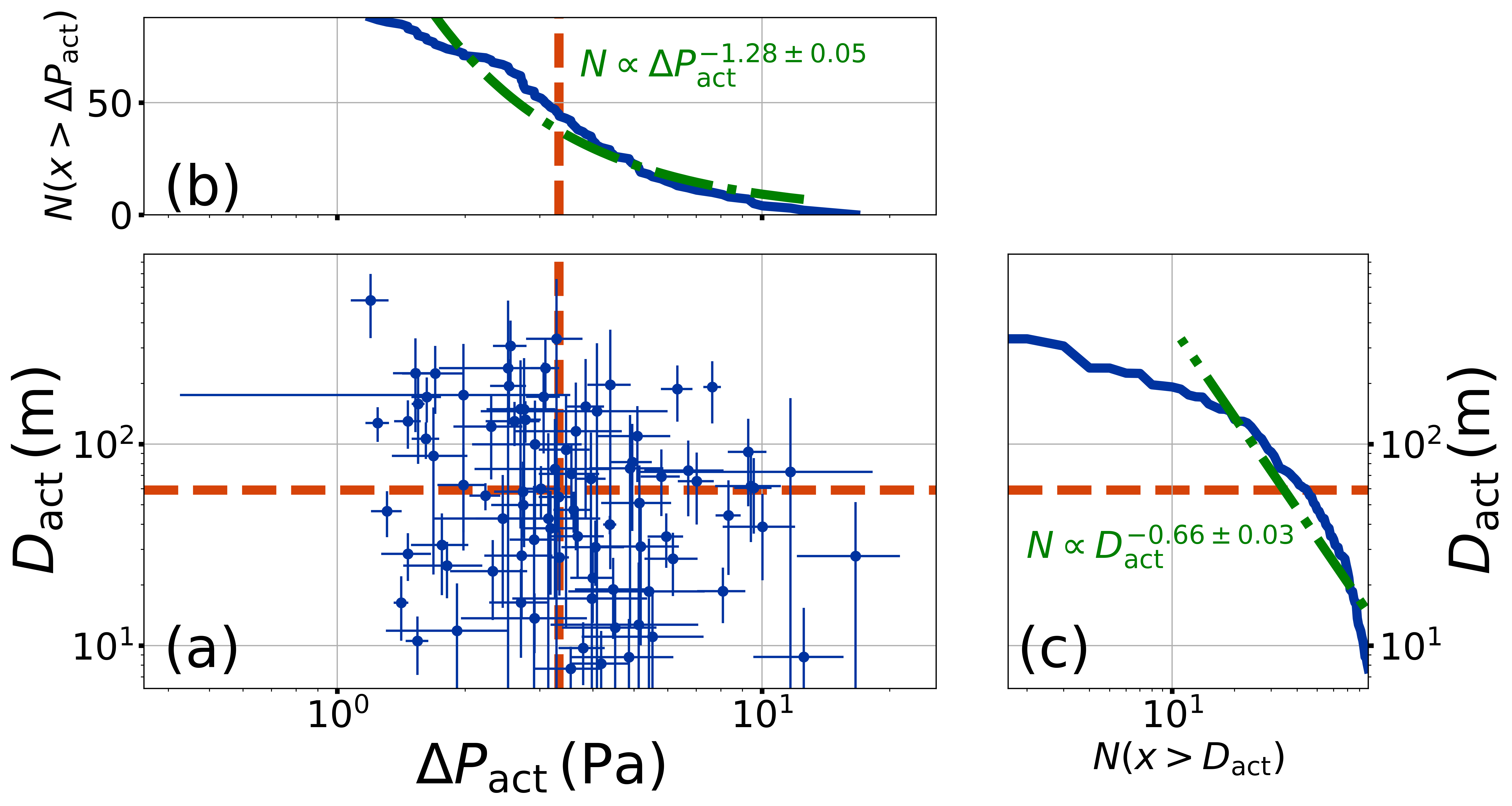}
    \caption{(a) Distribution of estimated actual vortex diameters $D_{\rm act}$ vs. the pressure excursions $\Delta P_{\rm act}$ (blue dots). (b) Cumulative histogram of $P_{\rm act}$-values. (c) Cumulative histogram of $D_{\rm act}$-values.}
    \label{fig:Dact_vs_Pact}
\end{figure}

The best-fit (\emph{i.e.}, minimum $\chi^2$) wind profile models are illustrated for the three representative vortex encounters in Figure \ref{fig:vortices_and_windspeed}. As discussed in Appendix \ref{sec:Inferring Encounter Geometries from the Pressure and Velocity Profiles}, $\Delta P_{\rm obs}$ and $V_{\rm obs}$, along with the assumption of cyclostrophic balance, gives $\Delta P_{\rm act}$ and $V_{\rm act}$. Figure \ref{fig:Dact_vs_Pact} shows the distribution of inferred $\Delta P_{\rm act}$- and $D_{\rm act}$-values. The minimum, median, and maximum values for $P_{\rm act}$ are $1.20\pm0.12$, $3.33\pm0.18$, and $16.6\pm4.5\,{\rm Pa}$, respectively, and for $D_{\rm act}$ are $7.70\pm2.19$, $59.1\pm17.6$, and $517\pm181\,{\rm m}$. Panels (b) and (c) show the cumulative histograms for $P_{\rm act}$ and $D_{\rm act}$, along with power-law fits. The fit for $P_{\rm act}$ appears to be significantly better than that for $D_{\rm act}$ and corresponds to a differential histogram with a power-law index $\gamma = -2.28$, significantly more shallow than the differential histogram for $P_{\rm obs}$ (with $\gamma = -3.39$). This result is qualitatively inconsistent with theoretical expectations that the distribution of actual pressure drops is steeper than (i.e., the magnitude of the power-law exponent is larger) or the same as the distribution of observed drops \citep{2014JAtS...71.4461L, 2018Icar..299..166J, 2019Icar..317..209K}.

This inconsistency likely arises from our choice to filter out distant encounters ($b/D_{\rm act} > 1$) and low-signal vortices ($V_{\rm obs}/\sigma < 3$). All else being equal, vortices with small $\Delta P_{\rm act}$ are also likely to have small $V_{\rm act}$ and therefore to register with small $V_{\rm obs}$ in an encounter. \citet{2020Icar..33813523J} also suggested that $D_{\rm act} \propto \Delta P_{\rm act}^{1/2}$, meaning small $\Delta P_{\rm act}$ are more likely to have $b/D_{\rm act} > 1$. The distribution of $D_{\rm act}$ vs.~$\Delta P_{\rm act}$ is statistically consistent with no correlation; however, a strict power-law fit gives $D_{\rm act} \propto \Delta P_{\rm act}^{-0.34}$. This unexpected (and possibly unrealistic) power-law index may arise from small-number statistics or our admittedly conservative choice to filter out distant and low windspeed encounters, rather than a true anti-correlation between the two parameters. (\citet{2019Icar..317..209K} discusses the expected relationship between $D_{\rm act}$ and $\Delta P_{\rm act}.$)

\begin{figure}
    \centering
    \includegraphics[width=0.5\textwidth]{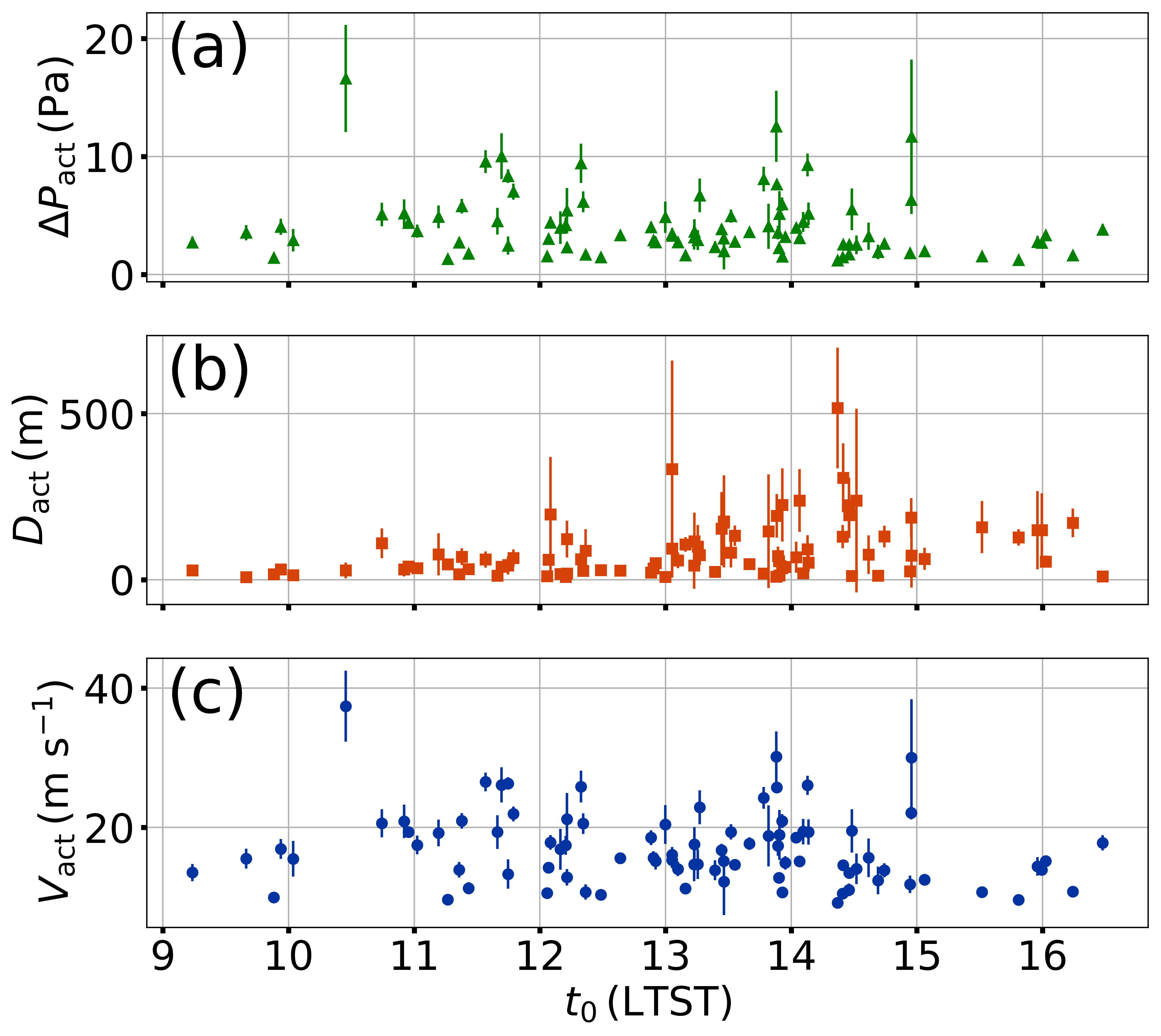}
    \caption{(a) Distributions of $\Delta P_{\rm act}$ (a), $D_{\rm act}$ (b), and $V_{\rm act}$ (c) with time-of-day $t_0$ (local true solar time, LTST).}
    \label{fig:all_actual_values_vs_t0}
\end{figure}

In spite of these observational biases, we can still infer eyewall velocities $V_{\rm act}$ and look at trends with time-of-day $t_0$. Figure \ref{fig:all_actual_values_vs_t0} shows how $\Delta P_{\rm act}$, $D_{\rm act}$, and $V_{\rm act}$ vary as functions of time-of-day $t_0$. The min, median, and maximum values for $V_{\rm act}$ are $9.17\pm0.48$, $15.6\pm1.0$, and $37.4\pm5.1\,{\rm m\ s^{-1}}$. The trends for all parameters with $t_0$ seem reasonable: many of the largest values for all parameters occur in the late afternoon, although the very largest values for each do not always occur late in the day. 

\begin{figure}
    \centering
    \includegraphics[width=0.5\textwidth]{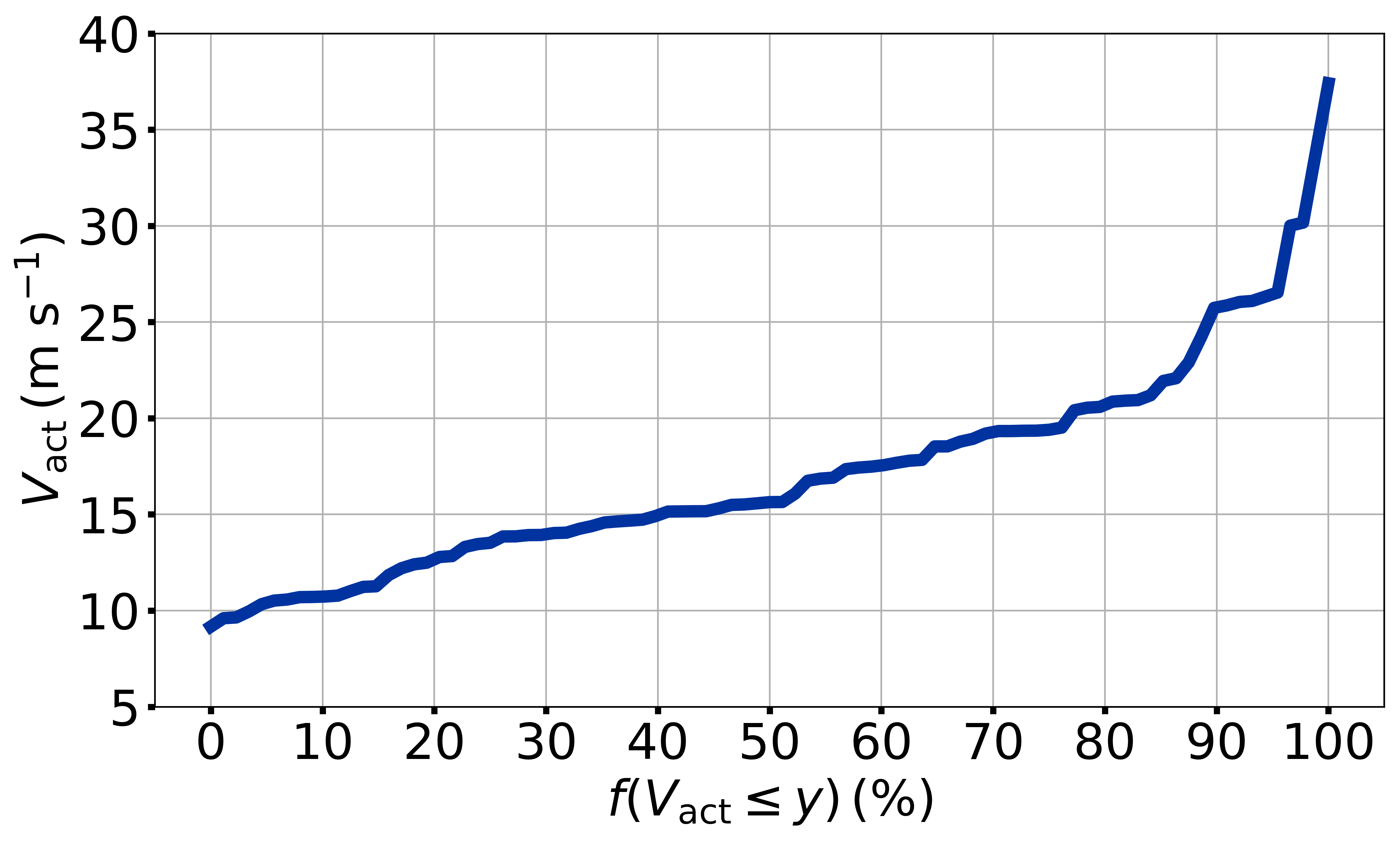}
    \caption{The fractional cumulative histogram of inferred $V_{\rm act}$-values. For example, about 20\% of the vortices we analyzed have $V_{\rm act} \geq 20\,{\rm m\ s^{-1}}$.}
    \label{fig:cum_hist_Vact}
\end{figure}

\subsection{Inferring Areal Occurrence Rate from the Time-Series Analysis}
\label{sec:Inferring Areal Occurrence Rate from the Time-Series Analysis}

\begin{figure}
    \centering
    \includegraphics[width=\textwidth]{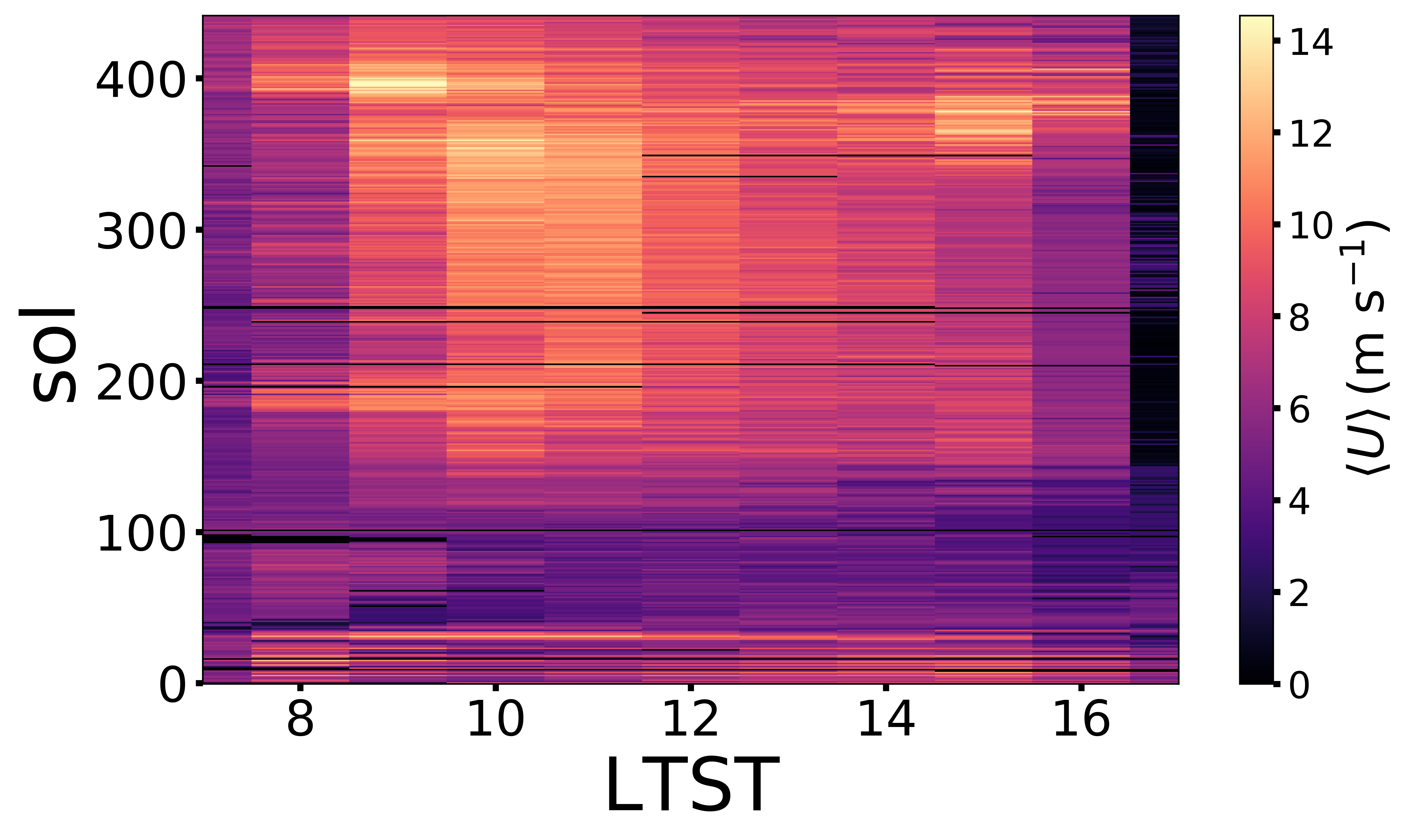}
    \caption{The hour-by-hour average wind speeds $\langle U \rangle$ for each sol. Brighter (yellower) colors represent higher speeds.}
    \label{fig:advective_speeds}
\end{figure}

If all encountered vortices had the same advective speed $U$, diameter $D_{\rm act}$, and central pressure excursion $\Delta P_{\rm act}$, the areal density of vortices $n$ can be estimated from the number of vortex detections per unit time, $\nu$, via
\begin{equation}
    \nu = k/T = n U \left( 2 b_{\rm max} \right) \label{eqn:simple_number_of_encounters}
\end{equation}

where $k$ is the total number of encounters during the observing period $T$ and $b_{\rm max} = \left( \frac{D_{\rm act}}{2} \right) \sqrt{ \frac{\Delta P_{\rm act}}{\Delta P_{\rm min}} - 1 }$ is the maximum radial encounter distance for which a pressure signal will register. $P_{\rm min}$ is the minimum observed pressure excursion for which a vortex encounter would register \citep{2021Icar..35814200K} and is taken as $0.1\,{\rm Pa}$ (see Section \ref{sec:Vortex Statistics}). For the advective speed, we use not the pre-encounter wind velocities considered in, for example, Equation \ref{eqn:total_wind_speed} but the hour-by-hour average speeds illustrated in Figure \ref{fig:advective_speeds} because we are interested in the advection of a population of vortices, not the individual vortices. In principle, calculating $n$ from the observed encounters requires integrating over the population. Unfortunately, the small number of vortices for which we were able to robustly estimate the actual parameters severely limits our ability to integrate over the population.  Moreover, we expect these parameters to vary with time-of-day and season. In lieu of this more complete evaluation, we instead calculate the population average for $b_{\rm max}$ and then use that average value to solve for $n$ from the encounter rates depicted in Figure \ref{fig:sol_and_t0_histograms}. We convert that areal density into an areal occurrence rate $f$ (number per area per time)  via the following equation:
\begin{equation}
    f = \left( \frac{k}{T} \right) \times \left( \dfrac{1}{2 b_{\rm max} T U } \right).\label{eqn:convert_to_areal_occurrence_rate}
\end{equation}
The quantity $T U$ represents the distance traveled by a vortex during the time observed, and $2 b_{\rm max}$ the width swept out by the vortex within which it would have been detectable.

\begin{figure}
    \centering
    \includegraphics[width=\textwidth]{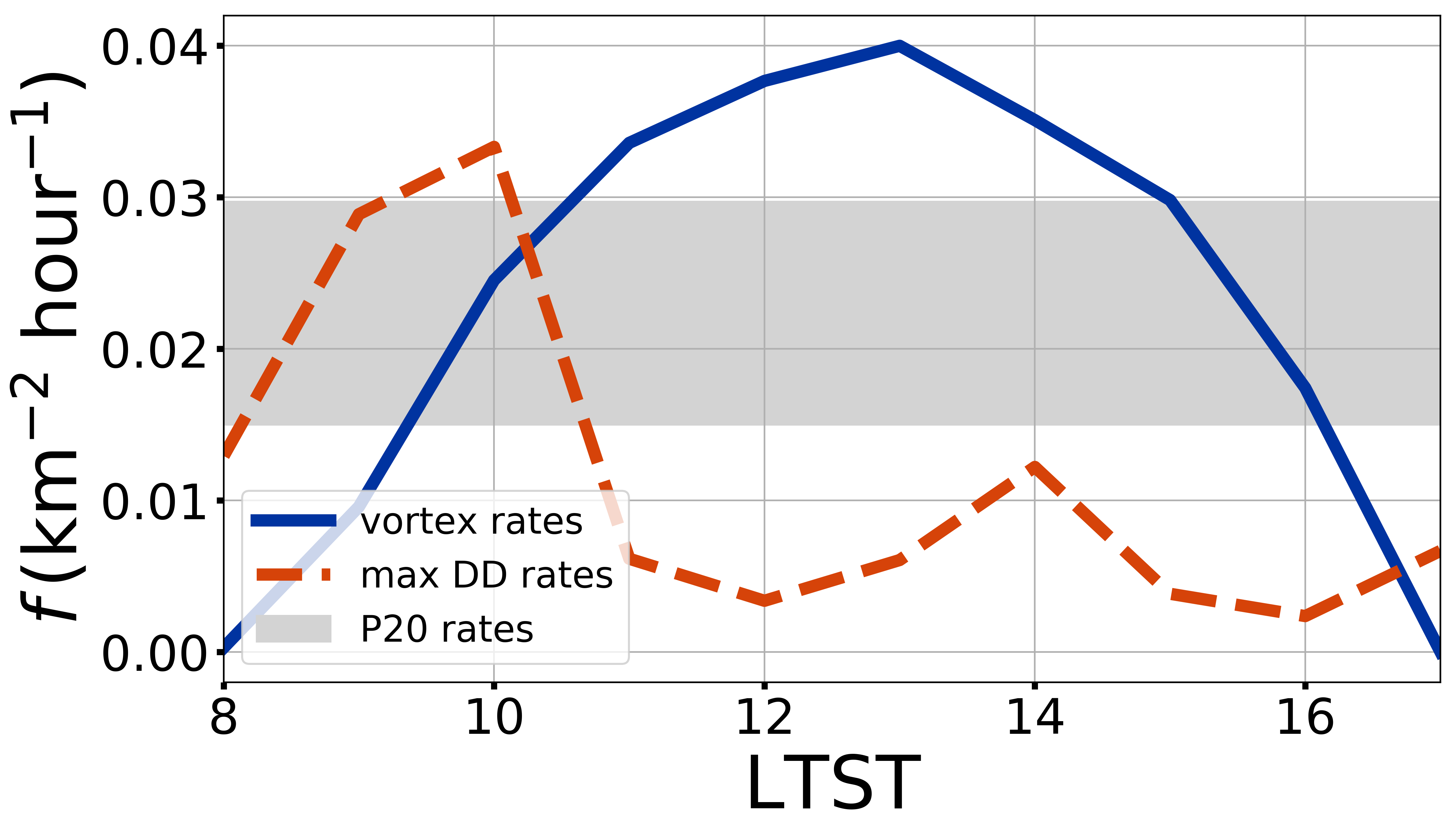}
    \caption{The solid blue curve shows the areal occurrence rate for vortices (``vortex rates''). The dashed, orange line shows the allowed maximum rate for dust devils inferred from the ICC image analysis (``max DD rates''). The grey region shows a range of rates reported in \citet{2016Icar..266..315R} and \citet{2020GeoRL..4787234P} (``P20 rates'').}
    \label{fig:areal_occurrence_rate}
\end{figure}

Figure \ref{fig:areal_occurrence_rate} shows our estimate for $f$ for vortices and compares it to values from other studies.

\section{Image Analysis}
\label{sec:Image Analysis}

\begin{figure}
    \centering
    \includegraphics[width=\textwidth]{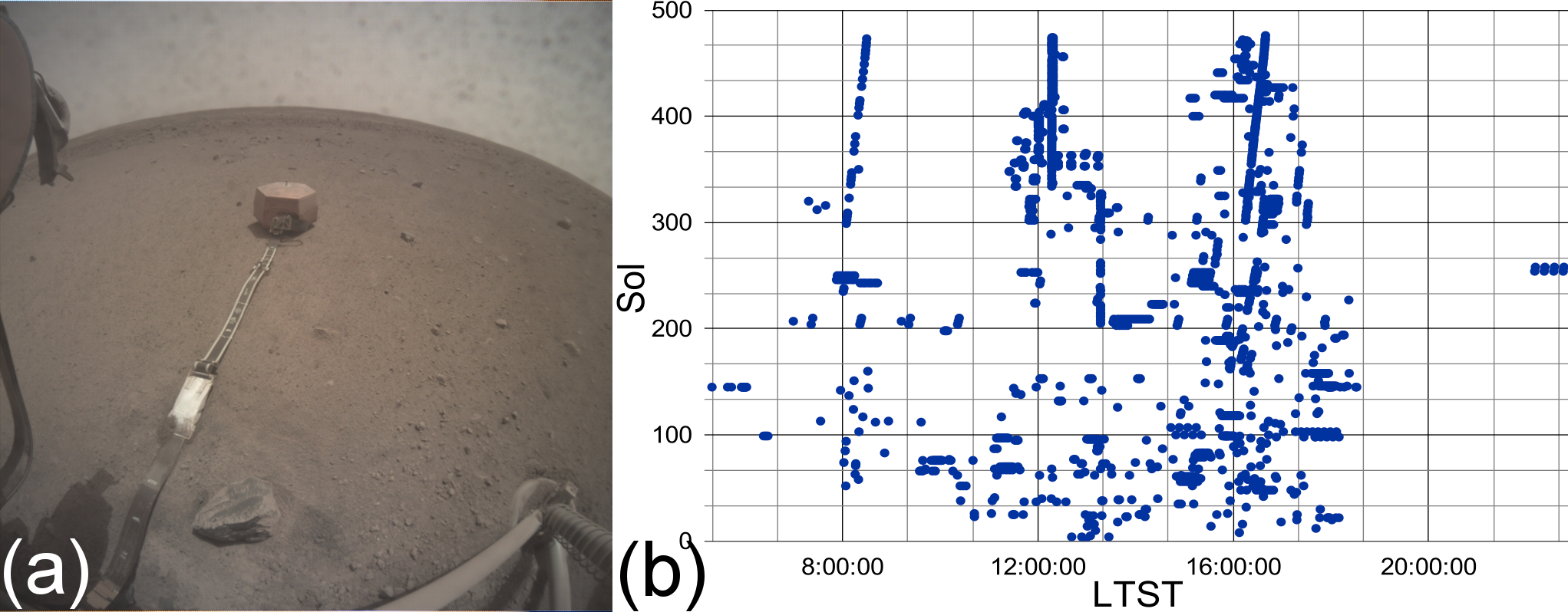}
    \caption{(a) Example of an Instrument Context Camera (ICC) image. (b) Times and sols during which images were collected throughout the InSight mission.}
    \label{fig:Example-Image_Insight-Combined-Analysis}
\end{figure}

In this section, we describe our analysis of image data from InSight to search for the passage of dust devils. To cut to the chase, our survey detected no active dust devils, but we can provide an upper limit on their activity based on the geometry, frequency, and timing of observations.

For our visual dust devil survey, we used observations from InSight's ICC because that camera is consistently pointed toward the horizon. ICC has a field-of-view (FOV) of $124^\circ\times124^\circ$ and sits about $0.7\,{\rm m}$ off the surface \citep{2018SSRv..214..105M} -- Figure \ref{fig:Example-Image_Insight-Combined-Analysis}(a) shows an example image. As of this manuscript's preparation, the NASA PDS archive for ICC contains images spanning sols 1 to 476 (although many sols lack images), totaling \numICCimages\ images. For our survey, three authors (JC, MS, RB) visually inspected all available ICC images hosted on the PDS archive (without any contrast enhancement or other aids). The inspections were conducted over the course of several weeks redundantly (i.e., multiple workers inspected the same image) and independently (to avoid biasing the results). Figure \ref{fig:Example-Image_Insight-Combined-Analysis}(b) shows the sols and times when images are available. 

Having seen no images of active dust devils, what upper limits can we place on their occurrence? The limits will depend on three aspects of the observations: (1) image contrast considerations (which constrain the maximum optical depth allowed for any unidentified dust devils), (2) the total area surveilled (which constrains the areal density of dust devils), and (3) the time span during which dust devils could have been spotted in each image (which allows us to convert the areal density to an areal occurrence rate).

Regarding image contrast, we invoke the procedure used in \citet{2006JGRE..11112S09G}. In that study, the authors analyzed images from the Mars Exploration Rover Spirit showing active dust devils and compared the values for pixels within dust devils $I_{\rm DD}$ to values for pixels showing the sky $I_{\rm sky}$ and the ground $I_{\rm ground}$. The study argued that a dust devil's optical depth $\tau$ could be estimated as

\begin{equation}
    \tau = \ln \left( \dfrac{I_{\rm ground} - I_{\rm sky}}{I_{\rm DD} - I_{\rm sky}}. \right)\label{eqn:optical_depth}
\end{equation}

We saw no active dust devils, and so we can use Equation \ref{eqn:optical_depth}, along with the distribution of pixel values from the ICC images, to estimate the maximum allowed optical depth for any dust-lofting vortices. By assumption, a dust devil (large enough to be resolved) could be spotted against the martian sky if it were considerably darker than the typical sky pixel. Inspecting some of the ICC images, we estimated median sky ($I_{\rm sky} \sim 150$) and ground ($I_{\rm ground} \sim 100$) pixel values, as well as the standard deviation for the sky pixels $\sigma_{I_{\rm sky}} \sim 15$. We assume a dust devil must be at least $3-\sigma_{I_{\rm sky}}$ darker than $I_{\rm sky}$ (i.e., $I_{\rm sky} - I_{\rm DD} \ge 3\ \sigma_{I_{\rm sky}}$) in order to be spotted. Using these values, we estimate the maximum optical depth for any dust devils hidden within the images as $\tau \lesssim 0.1$. This approach, of course, assumes a single pixel suffices to recognize a dust devil. However, a multi-pixel dust devil could be identified even with a smaller optical depth since more pixels would correspond to a higher signal-to-noise and therefore require a smaller contrast. Therefore, the estimate here represents a reasonable upper limit.

Regarding the area surveilled, the local topography limits the horizon for the Instrument Deployment Camera (IDC) toward the south to less than $2.4\,{\rm km}$ \citep{2020E&SS....701248G}. Since ICC is even closer to the ground ($0.7\,{\rm m}$ vs.~IDC's $1.5\,{\rm m}$), the view is even more limited. However, as an upper limit, we can estimate the area surveilled as $\frac{1}{2} \pi \left(124^\circ/180^\circ\right) \left( 2.4\,{\rm km} \right)^2 \approx 6.2\,{\rm km^2}$. An important limitation of this approach: smaller dust devils cannot be resolved at the same maximum distance as larger devils. However, ICC has an angular resolution of $\alpha = 2\times10^{-3}\,{\rm rad\ px^{-1}}$ \citep{2018SSRv..214..105M}, meaning we could resolve the diameter of the smallest vortex for which we could estimate a diameter ($D_{\rm act} = 7.7\,{\rm m}$) to a distance of about $3.9\,{\rm km}$, farther than the horizon.

Regarding the time span for an observation, the key timescale is the time for a dust devil to cross through the observational area, $T_{\rm cross}$. Given an areal occurrence rate $f$ and an observed area $A_{\rm obs}$, the total number of dust devils observed in one image would be $N = f A_{\rm obs} T_{\rm cross}$ (assuming the lifetime of the dust devils is long compared to $T_{\rm cross}$), which can be re-arranged to calculate $f$ given the other parameters. The maximum crossing time is equal to the time for a dust devil to cross along the horizon, no farther away than $2.4\,{\rm km}$. With an FOV of $124^\circ$, this distance is $\pi \left(124^\circ/180^\circ\right) \left( 2.4\,{\rm km} \right) \approx 2.6\,{\rm km}$. The time to cross this distance will depend on the ambient wind speed (with $U = 8\,{\rm m\ s^{-1}}$, $T_{\rm cross} \approx 5\,{\rm min}$), and so for our calculation, we take the median wind speed during each observational period. As an example, a single image showing no dust devils means $N < 1$, and therefore $f < \left( A_{\rm obs} T_{\rm cross}\right)^{-1} = \left( 6.2\,{\rm km^{2}}\ \times\ 5\,{\rm min}\right)^{-1} \approx 2\,{\rm km^{-2}\ hr^{-1}}$. Each additional image showing no dust devils reduces the allowed areal occurrence rate by another factor of $T_{\rm cross}$, assuming the images are separated in time by at least $T_{\rm cross}$. The upper limits on the areal occurrence rate incorporating the measured advection speeds and sol-by-sol images (Figure \ref{fig:Example-Image_Insight-Combined-Analysis}(b)) are shown in Figure \ref{fig:areal_occurrence_rate}. (Given that most sols have only one or a few images available, the areal occurrence rate binned by sol is much less informative, and so we do not include it.)

Folding all these considerations together with the number and timing of images reflected in Figure \ref{fig:Example-Image_Insight-Combined-Analysis}(b) and the advective wind speeds allows us to assess hourly upper limits on dust devil areal occurrence rates. Figure \ref{fig:areal_occurrence_rate} shows the result as a dashed orange line. To be clear, our null detection rules out dust devils within the ICC's field-of-view with $\tau > 0.1$ and subtending angles smaller than $2\times10^{-3}\,{\rm rad}$. Dust devils appearing within the available images with both a greater $\tau$ and a significantly larger angular diameter likely would have been spotted. 

These results comport with a recent study of the same dataset \citep{2021Icar..36414468L}. In that study, the authors conducted some injection-recovery experiments with the images and ruled out dust devils with optical depths greater than 1\% subtending an 8x16 pixel rectangle. The same study considered engineering data from Insight's solar panels to argue that most vortices were dustless, as we corroborate here.

A more comprehensive assessment based on the lack of imaged dust devils could provide a more detailed estimate of occurrence rates and optical depth. Such an assessment would likely require generation of images with (and without) synthetic dust devils, and an analogous survey of these images could then be conducted to assess detectability and robustness. However, such an exercise is beyond the scope of this study. We are interested only in upper limits, which our survey provides, implicitly including important effects such as image compression (any compression artifacts feed into the distribution of pixel values). We leave such a fuller assessment for future work.

\section{Discussion}
\label{sec:Discussion}

\subsection{Inferring Occurrence Rates and Thresholds for Dust Lifting}
\label{sec:Inferring Occurrence Rates and Thresholds for Dust Lifting}

Altogether, these results invite several interesting conclusions which are bolstered by and contrasted with previous studies. The lack of observed dust devils in the ICC imagery indicates that the vortices near InSight are frequently dustless and therefore invisible (at least to the limit of the image contrast). Figure \ref{fig:areal_occurrence_rate} shows the maximum areal occurrence rates for dust devils allowed by the image survey. Comparing to the rates of vortex occurrence suggests no more than 35\% of the encountered vortices could have lofted dust and still not have registered in the images. This rate appears roughly consistent with studies of terrestrial studies: deploying pressure loggers alongside solar sensors, \citet{LORENZ20151} found that 40\% of vortex events produced no solar attenuation, and only 20\% of events caused dimming greater than about 2\%. Studies on Mars have suggested martian vortices are very often dustless, especially when the boundary layer is shallow, which correlates with less vigorous vortices \citep{2015Icar..249..129M, 2016Icar..278..180S}. 

In fact, based on our non-detection of active dust devils and distribution of $V_{\rm act}$-values, we can estimate a minimum wind speed required to loft dust. If only 35\% of vortices lofted dust, Figure \ref{fig:cum_hist_Vact} suggests a threshold of $19\,{\rm m\ s^{-1}}$, corresponding to $\Delta P_{\rm act} \approx 7\,{\rm Pa}$. (N.B. $\Delta P_{\rm act} \geq \Delta P_{\rm obs}$.) Of course, this value is a minimum since an even smaller fraction of dusty vortices would still be consistent with our null detection among the ICC images, but it appears roughly consistent with other work. Based on lab simulations, \citet{2003JGRE..108.5041G} proposed $20-30\,{\rm m\ s^{-1}}$. Results from a study tracking the motions of dust clots within martian dust devils agreed with that estimate \citep{2011GeoRL..3824206C}. Other studies suggest much higher thresholds \citep[\emph{cf.}][]{2006JGRE..11112002C}. We can, of course, also flip the problem around and assume a minimum wind speed for lofting dust and then use Figure \ref{fig:cum_hist_Vact} to infer the fraction of vortices that would be expected to be dust devils.

We can also consider thresholds required to form tracks on the martian surface. As a vortex travels over the surface, it may disrupt the surficial sediment, revealing a brighter or darker surface beneath. Previous studies have used observations either in-situ or from orbit of dust devil tracks to infer the diameters, lifetimes, and occurrence rates of dust devils \citep[\emph{e.g.},][]{2008JGRE..113.7002W}. (However, a vortex may leave a track without lofting significant dust, and not all dust devils leave discernible tracks -- \citealp{2005JGRE..110.6002G}.) 

With these caveats in mind, we can once again compare areal occurrence rates to estimate the fractions of vortices and dust devils leaving tracks. \citet{2016Icar..266..315R} and \citet{2020GeoRL..4787234P} both conducted surveys of the region surrounding InSight for dust devil tracks, and Figure \ref{fig:areal_occurrence_rate} shows the range of rates from those studies. (We have excluded the very large rate of $0.68\,{\rm km^{-2}\ sol^{-1}}$ reported in \citet{2020GeoRL..4787234P} as an outlier.) The rates are reported in ${\rm km^{-2}\ sol^{-1}}$, so we multiplied them by 9/24 to convert from per sol to per hour. (Our vortex encounter analysis indicated that vortices are active for about 9 martian hours for each martian sol -- Figure \ref{fig:sol_and_t0_histograms}b.) Comparison to our inferred occurrence rates suggests between 38 and 74\% of vortices leave tracks. This result may mean that between 26 and 62\% have insufficient wind speeds to visibly rearrange the surficial sediment. Looking again at Figure \ref{fig:cum_hist_Vact}, these numbers correspond to $V_{\rm act} = 14$ and $18\,{\rm m\ s^{-1}}$, respectively. 

We can also gauge how often a dust devil might leave a trail (at least, in the region surrounding InSight). Around 14:00 LTST, the largest allowed dust devil occurrence rate is $0.01\,{\rm km^{-2}\ hour^{-1}}$. This rate is about 80\% of the minimum track formation rate. As with the comparison to the ICC images, vortex diameter may factor into these considerations: vortices may be too small to leave tracks resolvable by the HiRISE instrument. However, \citet{2020GeoRL..4787234P} reported a few tracks with widths as small as $5\,{\rm m}$ (but none smaller), indicating that, even our smallest vortex with $D_{\rm act} = 7.7\,{\rm m}$ could have left a resolvable track. Work on the track dataset continues to determine the distribution of diameters, and so further comparisons must wait.

\subsection{Comparison to Previous Work}
\label{sec:Comparison to Previous Work}

Our results corroborate some results from previous studies. \citet{2021Icar..35514119L} conducted a survey of InSight meteorological data very similar to this one, recovering 853 events with pressure excursions exceeding $0.8\,{\rm Pa}$ over the first 390 sols of the InSight mission and amounting to 2-3 encounters per sol, similar to our encounter rates (Figure \ref{fig:sol_and_t0_histograms}b). Although that study analyzed the pressure profiles of individual vortex encounters and the ambient wind speeds adjacent to an encounter, it did not model the vortex wind profiles. That study did consider the seismic signals from vortex passes as well. Unfortunately, \citet{2021Icar..35514119L} did not explicitly estimate an areal occurrence rate for vortices for direct comparison to our results but instead the fractional surface area occulted by vortices $F$. This parameter was estimated by dividing the total duration of encounters to the total duration of data collections during the hours when vortices are active, giving $F \approx 0.07\%$. Performing the same calculation for our detections, we find $F \approx 0.08\%$, indicating good agreement between our results. Comparing the two catalogs, we see that \citet{2021Icar..35514119L} found 75\% of our encounters from before sol 390.

\citet{Spiga2021}'s survey also resembles this survey but reported a much higher vortex encounter rate: 6046 in the mission's first 400 sols. Comparing detection catalogs, we find that study recovered nearly 90\% of our detections, indicating good agreement for the encounters we found. The reasons that \citet{Spiga2021} found so many more encounters are not entirely clear but probably arise from our different detection schemes. \citet{Spiga2021} fit a straight-line to 1000-second windows surrounding each data point in the pressure time-series and then recorded any negative excursions greater than $0.35\,{\rm Pa}$ as encounters. This approach might record any negative pressure excursion, regardless of its duration or time structure, as a vortex encounter, while our approach might filter out some signals that are not sufficiently Lorentz-like. Following a statistical approach similar to ours, \citet{2021Icar..35814200K} determined that the detections in \citet{Spiga2021} imply a vortex occurrence rate of 56 vortices per ${\rm km}^2$, which was described as ``an unprecedented high level''.

It is plausible that some of the disagreement between studies arises from the different detection thresholds. \citet{2021Icar..35514119L} required $\Delta P_{\rm obs} > 0.8\,{\rm Pa}$, while \citet{Spiga2021} required $\Delta P_{\rm obs} > 0.35\,{\rm Pa}$. (Our detection threshold is not quantified in the same way -- see Appendix \ref{sec:Vortex Recovery Statistics}.) For the cumulative histogram of $\Delta P_{\rm obs}$-values, \citet{2021Icar..35514119L} inferred a power-law with an index of about $-2$, nearly consistent with ours (i.e., the number of encounters with a $\Delta P_{\rm obs}$-value or higher drops as $\Delta P_{\rm obs}^{-2}$). \citet{2021Icar..35514119L} reported 853 detections and so would have expected 4460 ($= 853 \times (0.8\,{\rm Pa}/0.35\,{\rm Pa})^2$) total detections with $\Delta P_{\rm obs} \geq 0.35\,{\rm Pa}$, inconsistent with the 6000 detections of \citet{Spiga2021}. \citet{Spiga2021} inferred a similar dependence for the $\Delta P_{\rm obs}$ cumulative histogram, reporting a power-law index between -2.5 for the smallest-$\Delta P_{\rm obs}$ encounters. Taking the former index and their total number of detections (6046), \citet{Spiga2021} would have expected 765 ($= 6046 \times \left(0.35\,{\rm Pa}/0.8\,{\rm Pa} \right)^{2.5}$) total detections with $\Delta P_{\rm obs} \geq 0.8\,{\rm Pa}$, not entirely consistent with the 853 detections of \citet{2021Icar..35514119L} -- Poisson statistics suggests disagreement at more than 3-$\sigma$ ($\sqrt{853} \approx 30$). Given their good agreement with \citet{2021Icar..35514119L}, our results also appear inconsistent with those of \citet{Spiga2021}.

The overall encounter rate from \citet{Spiga2021} is about seven times larger than our rate and therefore implies an areal occurrence rate of about $0.28\,{\rm km^{-2}\ hour^{-1}}$. This result suggests no more than 1 out of about 28 vortices (0.28/0.01) is a visually detectable dust devil, which might suggest martian vortices are much less likely to loft dust than terrestrial ones \citep{LORENZ20151}. This areal occurrence rate is also ten times larger than the rate for tracks, $\le 0.03\,{\rm km^{-2}\ hour^{-1}}$.

As measured in the average number of vortices encountered per sol, our results suggest vortices are active at InSight at a level comparable to sites for previous missions. \citet{2010JGRE..115.0E16E} reported 502 vortex encounters by the Phoenix mission, which landed at $68.2^\circ$ N, over 151 sols from $L_{\rm s} = 76^\circ$ to $148^\circ$. The lander encountered about 3 vortices per sol, with seasonally varying mid-day peaks from about 0.2 to 0.8 per hour, rates only slightly larger than the rates we report here. The total duration of encounters normalized to the total observational time suggests a fractional area occulted $F \approx 0.01\%$, smaller than the fractional area estimates in our study or \citet{2021Icar..35514119L}. \citet{2010JGRE..115.0E16E} also conducted an imaging survey at the Phoenix site, imaging 37 unique dust devils; however, the requisite data to convert those detections into an areal occurrence rate are not provided. (An unspecified number of devils are imaged multiple times.) \citet{2019JGRE..124.3442N} identified vortex encounters using three martian years of pressure data from the Mars Science Laboratory (MSL) in the vicinity of Gale Crater, near $5.3^\circ$ S. That study found similar per-sol and per-hour encounter rates to what we report here. Detailed atmospheric modeling allowed a comparison between observed encounter rates and the expected meteorological conditions, corroborating some theoretical expectations \citep{1998JAtS...55.3244R}. \citet{2020Icar..34713814O} conducted a survey of MSL including data from beyond the mission's third year and found a significant increase in dust devil activity which was attributed to higher elevation of the terrain, lower thermal inertia of the environment, and more available dust.

However, cast in terms of the number of vortices with a given $\Delta P_{\rm obs}$-value, the InSight landing site does appear to be significantly more active than the Phoenix site, as suggested by \citet{Spiga2021} and \citet{2021Icar..35514119L}. Extrapolating our $\Delta P_{\rm obs}$ power-law fit (Figure \ref{fig:DeltaPobs_vs_Gammaobs}) and the per-sol number of encounters, we might have expected more than 7,000 encounters with $\Delta P_{\rm obs} > 0.3\,{\rm Pa}$. The power-law fits from \citet{Spiga2021} and \citet{2021Icar..35514119L} give different expectations, but all agree that the number of encounters actually reported is significantly less than expected based on the InSight encounters.

The dust devil areal occurrence rate inferred for the Spirit lander site in \citet{2010JGRE..115.0F02G} appears to be comparable to the rate for vortices we infer for InSight. \citet{2010JGRE..115.0F02G} analyzed images collected over three martian years, netting more than 700 sightings of active dust devils. Normalizing their detections by the imaging area and frequency, they inferred hourly areal occurrence rates which varied over a sol up to $0.05\,{\rm km^{-2}\ hour^{-1}}$ (their Figure 7). Of course, theirs was an imaging survey and ours is a meteorological survey.

\subsection{Why Didn't InSight Image Any Dust Devils?}
\label{sec:Why Didn't InSight Image Any Dust Devils?}

If the vortex occurrence rate at InSight is similar to or even greatly exceeds the rates seen by other Mars landers, why did InSight image no dust devils when those other landers imaged many? Not for lack of trying: \citet{2020NatGe..13..190B} describe a concerted imaging campaign to detect dust devils, as illustrated in Figure \ref{fig:Example-Image_Insight-Combined-Analysis}(b). This imaging campaign resembles campaigns conducted by those other missions, with more than a thousand images collected over hundreds of sols. 

\begin{figure}
    \centering
    \includegraphics[width=\textwidth]{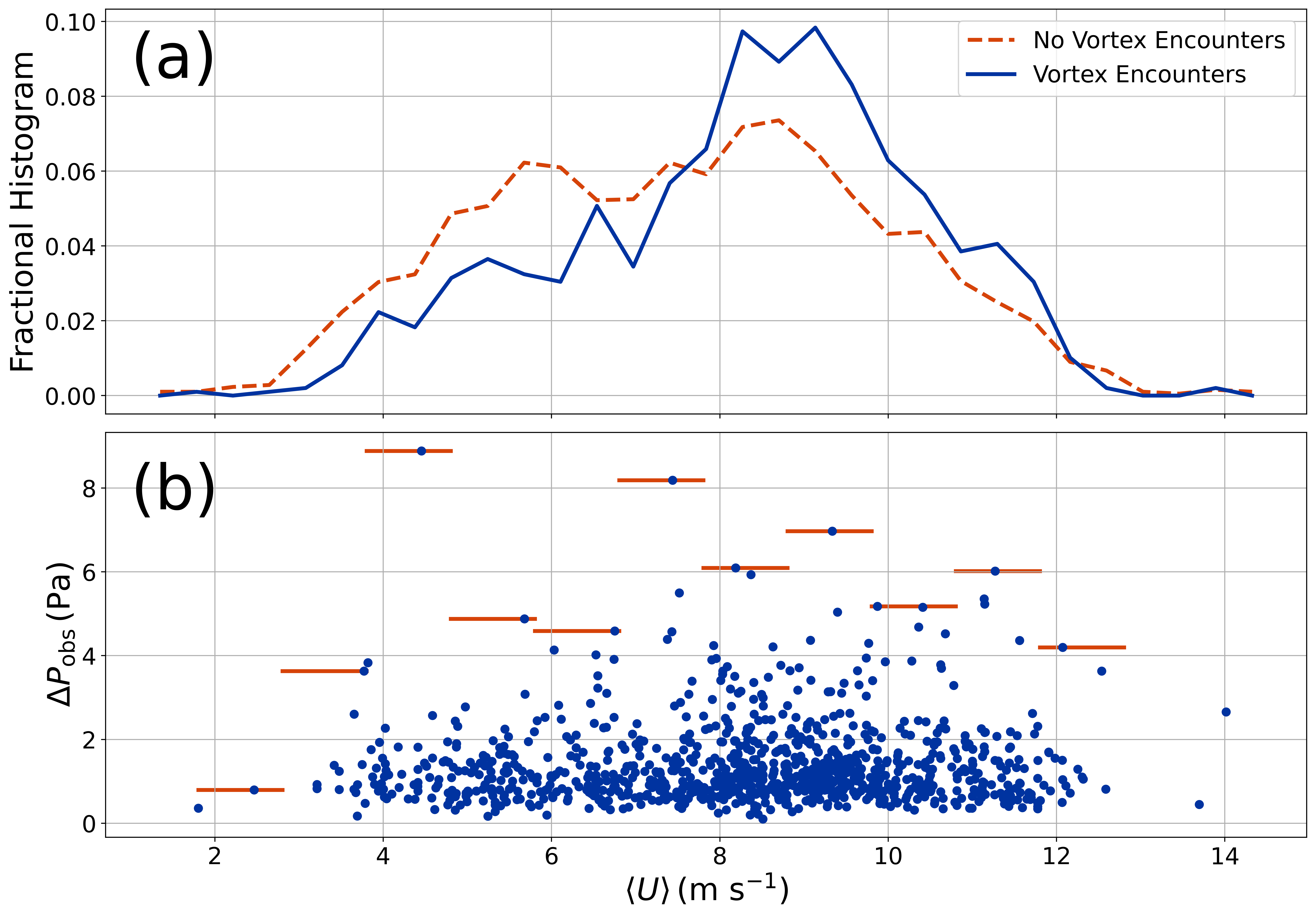}
    \caption{(a) The distribution of hourly-averaged wind speeds during the hours when vortices were encountered (solid, blue line) and between 8:00 and 16:00 LTST during hours when they were \emph{not} encountered (dashed, orange line). (b) Vortex $\Delta P_{\rm obs}$-values vs.~their hourly averaged advective wind speeds. The horizontal orange lines show the maximum $\Delta P_{\rm obs}$-value observed for wind speeds binned by $2\, {\rm m\ s^{-1}}$ (the horizontal span of each line shows the binning).}
    \label{fig:U_vs_DeltaP_comparisons}
\end{figure}

Inspection of the hourly wind speed data collected throughout the mission suggests one explanation for the lack of imaged dust devils: the InSight landing site appears to be much windier during the times of day when vortex activity occurs. Figure \ref{fig:U_vs_DeltaP_comparisons}(a) shows the distribution of hourly-averaged wind speeds both during vortex encounters and between 8:00 and 16:00 LTST but during hours when no vortices were encountered. (N.B., these wind speeds are different from the speeds from immediately before encounters used to estimate vortex diameters and shown in Figure \ref{fig:U1_vs_Gamma_hist}). Clearly, the vortex-associated advective speeds skew toward larger values than the winds overall, with an average of $8.3\,{\rm m\ s^{-1}}$. 

The seminal terrestrial field studies of \citet{1969JApMe...8...32S} indicated that dust devil frequency often increases for increasing wind speed but then declines again above a certain wind speed. The same trend seems to hold for martian dust devils. Among imaged dust devils for which horizontal speeds could be estimated, \citet{2010JGRE..115.0F02G} found only about two dozen of about 500 total advected faster than $8\,{\rm m\ s^{-1}}$. Though \citet{2010JGRE..115.0E16E} did not directly estimate the advective velocities of imaged dust devils, the hourly-averaged wind speeds measured by Phoenix rarely exceeded $8\, {\rm m\ s^{-1}}$.

Certainly, an increased advection speed would be expected to increase the rate of dust devil encounter since more dust devils would be advected past the camera or meteorological sensor, what has been called the ``advection effect''. Moreover, some non-zero winds are probably necessary to provide the vorticity requisite for dust devil formation \citep{2020Icar..33813523J}, and, in any case, turbulent winds must accompany the convectively unstable conditions that produce dust devils. However, as discussed in \citet{2016SSRv..203..183R}, higher wind speeds may suppress a high near-surface lapse rate and reduce the vigor of convective mixing associated with dust devils. In addition, wind shear could disrupt the dynamical structures in which dust devils are embedded.

Evidence for the suppression of dust devils at high wind speeds appears in Figure \ref{fig:U_vs_DeltaP_comparisons}(a): there are fewer vortex encounters above about $9\,{\rm m\ s^{-1}}$. Moreover, Figure \ref{fig:U_vs_DeltaP_comparisons} shows the maximum $\Delta P_{\rm obs}$ for vortices that \emph{are} encountered seems to decline for $\langle U \rangle$ exceeding $4\,{\rm m\ s^{-1}}$, qualitatively consistent with the suggestion that high wind speeds disrupt the structure and therefore vortex strength. It is worth noting that the median $\Delta P_{\rm obs}$-values remain roughly constant with $\langle U \rangle$; however, all else equal, we might reasonably expect it is the most vigorous vortices which are dust devils, not necessarily the average vortices. Also, although the number of vortex encounters declines as $\langle U \rangle$ passes $8\, {\rm m\ s^{-1}}$, thereby potentially reducing the width of the $\Delta P_{\rm obs}$ distribution, the systematic decline in maximum $\Delta P_{\rm obs}$ with increasing $\langle U \rangle$ appears well before the maximum in number of vortices binned by $\langle U \rangle$.

Ostensibly, these results seem to contradict some of those from \cite{Spiga2021}, who reported a strong positive correlation between wind speed and vortex encounter rate and reasonably invoked the advection effect. That study also explored the dependence of vortex occurrence on meteorological conditions using large eddy simulations (LES) and found an increase in encounter rate with the synthetic vortices from the model but a reduction in the vortex areal occurrence rate. (Thus, they attributed their increased encounter rate to the advection effect.)
However, the wind speeds they reported for vortex encounters were not the hour-by-hour averages we discuss here. Instead, they calculated the average between 11:00 and 14:00 LTST for each day, which gave smaller speeds. In fact, if we conduct the same wind speed averaging, we obtain similar results (smaller overall average wind speeds and a montonic increase in encounter rate with wind speed). 

Why the InSight landing site might have been windier overall than the landing sites for other mission that successfully imaged dust devils is not clear and should be the focus of future studies. Moreover, the comprehensive data and/or studies for other Mars missions are lacking to robustly corroborate the possibility that InSight's landing site is indeed windier than other sites. On top of all that, multiple cases may contribute to suppression of dust devils at InSight. Although numerous nearby dust devil tracks have been imaged from space, it is possible that there is also a lack of liftable dust. In any case, the available data are at least consistent with the idea that dust devil formation was suppressed at InSight by higher wind speeds than have been seen at other landing sites. 

\section{Conclusions}
\label{sec:Conclusions}

Our analysis of InSight's APSS pressure and wind speed time-series search has netted 990 encounters with low-pressure, high-wind vortices over the first 477 sols of the mission, on average two encounters per sol (Figure \ref{fig:sol_and_t0_histograms}), in good agreement with some previous studies of InSight data \citep{2021Icar..35514119L} and somewhat inconsistent with others \citep{Spiga2021}. The distribution of observed pressure excursions associated with these vortices also resembles the distributions from other martian vortex studies (Figure \ref{fig:DeltaPobs_vs_Gammaobs}). 

Our analysis of wind speeds from the TWINS instrument allowed us both to infer the advection speeds for the vortices and vortex wind profiles themselves. These advection speeds allowed us to convert encounter rates into the intrinsic vortex occurrence rates (solid, blue line in Figure \ref{fig:areal_occurrence_rate}), and we found reasonable agreement with previous meteorological studies of both the InSight and other lander sites (Section \ref{sec:Discussion}). By leveraging assumptions about the pressure and wind profile shapes similar to previous work \citep{2016Icar..271..326L}, we were able to estimate the encounter distances between the vortex centers and the InSight lander for many of the vortices and back out the maximum wind velocities. Assuming a minimum threshold for dust lifting of about $20\,{\rm m\ s^{-1}}$, we estimated that about 35\% of encountered vortices would have been bonafide dust devils. This result agrees with terrestrial field studies about how often vortices may loft visible dust \citep{LORENZ20151}.

We also surveyed 1577 images (Figure \ref{fig:Example-Image_Insight-Combined-Analysis}) collected by InSight's ICC. Seeing no active dust devils, we were able to put an upper limit on the fraction of vortices that lift significant amounts of dust (dashed, orange line in Figure \ref{fig:areal_occurrence_rate}), no more than 35\% of vortices. It is crucial to note that this value is an upper limit with many limitations and caveats. Future work may revise this result. In any case, comparison of the distribution of wind velocities and the occurrence rates to results from studies of dust devil tracks seen from orbit in the region around InSight \citep{2020GeoRL..4787234P} allowed us also to infer that probably not all track-forming vortices are dust devils, and perhaps no more than 74\% of vortices leave tracks. Consequently, assuming the tangential wind speed is the only important factor in track formation, the minimum speed required may be $14\,{\rm m\ s^{-1}}$ (Section \ref{sec:Discussion}). By exploring the relationships between vortex encounters and parameters with advective wind speeds, we also found evidence that the lack of dust devils imaged by InSight might arise from high wind speeds, although multiple causes may contribute. In addition, we do not suggest high wind speeds at InSight suppressed vortex formation generally, just formation of the most vigorous vortices, the vortices most likely to be dust devils.

As impactful as these results may be, they involve a number of important assumptions and limitations. Perhaps most important, the turbulence of winds at the martian surface introduced considerable correlated noise \citep[\emph{cf.}][]{2018RemS...10...65J} into the TWINS wind speed, frequently obscuring the wind profiles of encountered vortices. Consequently, we limited our inference of vortex encounters to those with encounter distances less than one vortex diameter. This approach limited the number of encounters for which we could estimate intrinsic parameters and probably biased our inferred wind speed distribution to only the largest and/or most vigorous vortices. Fortunately, time-analysis techniques to account for such non-white noise exist \citep{hodlr} and should be considered in future work. 

The relatively slow sampling rate for TWINS ($1\,{\rm Hz}$) also presented issues. Since vortex encounters often last for only a few seconds (Figures \ref{fig:vortices_and_windspeed} and \ref{fig:DeltaPobs_vs_Gammaobs}), such sampling often only provided a few points during the encounter, challenging robust inference of the profile parameters. Of course, data volume is always an issue with planetary missions, but perhaps future missions that include meteorological instrumentation could consider short, high-resolution monitoring campaigns to more accurately capture vortex behavior during times of sol when they are expected to be most active. Sampling of $10\,{\rm Hz}$ or better would also allow more accurate assessment of other important boundary layer processes, such as turbulent heat and momentum transport \citep{2011RvGeo..49.3005P}.

Standards for reporting vortex analyses and statistics would also significantly facilitate comparison between studies of the same and of different datasets. Such comparisons not only help corroborate results from different studies but could also make more robust possible detections of time-variability in vortex and boundary layer behavior. For instance, in lieu of publishing only summary statistics and histograms of vortex properties, authors should consider providing tables of the detections themselves, including links to, for example, the specific images in which dust devils were detected.

In the future, planetary missions with a focus on or at least capabilities to assess boundary layer phenomena will elucidate these important and crucial processes. Since all surface-atmosphere interactions are mediated through these processes, they play key roles in shaping not just the climate but also the geology of worlds throughout the solar system, even on small bodies with only the barest breath of an atmosphere \citep{2017PNAS..114.2509J}. Fortunately, the increasing number of active and future missions carrying meteorological equipment bodes well for studies of surface-atmosphere interactions and, in particular, convective vortices and dust devils. 

\acknowledgments

We acknowledge helpful input from Don Banfield, Matthew Golombek, Ralph Lorenz, Patrick Whelley and two anonymous referees. We also thank the InSight team and NASA PDS for providing access to the data. The data are available from NASA's Atmosphere's PDS Node - \url{https://atmos.nmsu.edu/data_and_services/atmospheres_data/INSIGHT/insight.html}. BJ was supported by a grant from NASA's Solar System Workings program NNH17ZDA001N. JC, MS, and RB were supported by a grant from the Idaho Space Grant Consortium. All results from this study including the analysis codes are available here - \url{https://github.com/BoiseStatePlanetary/Recovering-Martian-Dust-Devil-Population}.

\vspace{5mm}

\software{matplotlib \citep{Hunter:2007}, numpy \citep{harris2020array}, scipy \citep{2020SciPy-NMeth}, statsmodels \citep{seabold2010statsmodels}}

\appendix

\section{Vortex Recovery Statistics}
\label{sec:Vortex Recovery Statistics}
In this section, we describe our analysis of our vortex recovery statistics. We explored the effects of the mean boxcar filter on both the time-series scatter but also on the detected vortices, as well as the effectiveness of our matched filter approach.

\begin{figure}
    \centering
    \includegraphics[width=\textwidth]{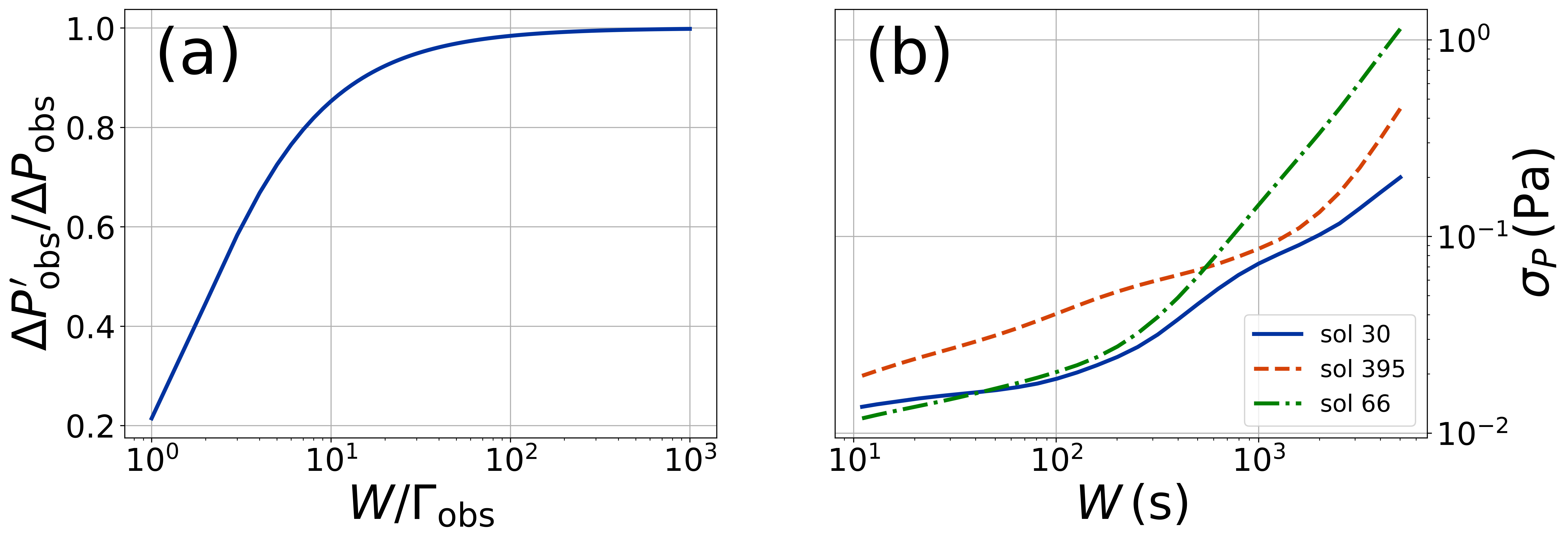}
    \caption{(a) Comparing the pressure excursion observed before $\Delta P_{\rm obs}$ application of the mean boxcar filter and after $\Delta P_{\rm obs}^\prime$ as a function of the width of the filter $W$ and of the vortex signal $\Gamma_{\rm obs}$. (b) Scatter in the pressure time-series $\sigma_P$ for the sol with the largest (sol 435) and least (sol 269) values as a function of the window size $W$ for the mean boxcar filter. Sol 395 has scatter close the median value for all sols. }
    \label{fig:Pobsprime-sigmaP_vs_W}
\end{figure}

For our study here, the mean boxcar filter  acts as high-pass filter on the APSS pressure time-series and, in principle, should induce little distortion on signals much narrower than the filter window size $W$. However, some vortices have quite long durations (many tens of seconds), and so they may be distorted if we use a small enough window. As a measure of this distortion, we can calculate how much less deep a vortex profile would appear after applying the filter by calculating the convolution of a boxcar function against a Lorentzian profile:
\begin{equation}
    \Delta P_{\rm obs}^\prime = \int_{t = -W/2}^{+W/2} \left( \frac{1}{W} \right) \left( \frac{-\Delta P_{\rm obs}}{1 + \left( \frac{t}{\Gamma_{\rm obs}/2} \right)^2} \right) dt = -\left( \frac{\Delta P_{\rm obs} \Gamma_{\rm obs}}{W}\right) \tan^{-1} \left( \frac{W}{\Gamma_{\rm obs}} \right), \label{eqn:tophat_convolution}
\end{equation}
where we have taken the Lorenztian to be centered at $t_0 = 0$ and $\Delta P_{\rm obs}^\prime$ represents the distorted profile depth. Figure \ref{fig:Pobsprime-sigmaP_vs_W}(a) shows the result: for windows more than 100 times the profile's original width (i.e., $W/\Gamma_{\rm obs} > 10^2$), the profile depth is more than 98\% of its original value, indicating minimal distortion. 

Of course, the narrower the window, the more effectively we can reduce the long-term variations in the time-series that may otherwise obscure the vortices. To explore that effect, we applied mean boxcar filters of various widths $W$ to each sol's pressure time-series and then estimated the resulting scatter (via $1.4826\ \times$ the median absolute deviation \citealp{doi:10.1080/01621459.1993.10476408}). Figure \ref{fig:Pobsprime-sigmaP_vs_W}(b) shows that for the time-series for sol 66 exhibited the largest scatter for any value of $W$, while that for sol 30 exhibited the smallest. The time-series for sol 395 had values near the median for all sols. In all cases, as $W$ increases, so does the scatter, consistent with our expectations that less aggressive filtering (i.e., $W$ larger) leaves more noise. We fit Lorentzian profiles to the vortices reported in \citet{Spiga2021} and found that the largest $\Gamma_{\rm obs} \approx 300\,{\rm s}$. Therefore, we took $W = 3000\,{\rm s}$, meaning that even the most distorted vortices should have $\Delta P_{\rm obs}^\prime/\Delta P_{\rm obs} > 0.8$.

\begin{figure}
    \centering
    \includegraphics[width=\textwidth]{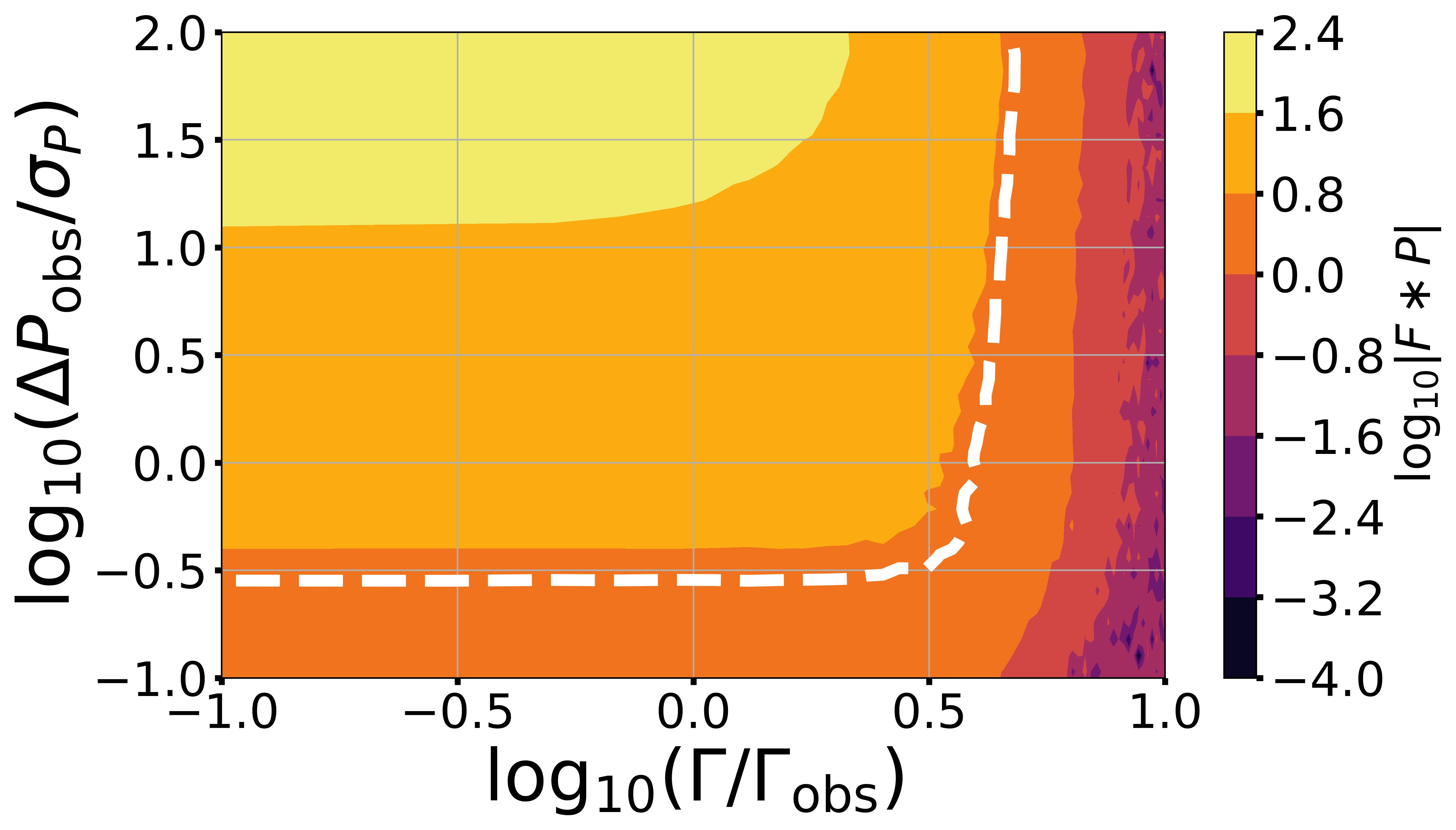}
    \caption{How effectively a convolution of a synthetic pressure time-series with the matched filter ($F \ast P$) recovers a vortex. The vortex has a known depth $\Delta P_\text{obs}$ and width $\Gamma_\text{obs}$ and is embedded in a synthetic time-series with white noise of variance $\sigma_P$. The matched filtered has a width $\Gamma$. The dashed, white line shows the threshold for detection used in this study ($F \ast P \geq 5$). In principle, vortices with values of $\Delta P_{\rm obs}/\sigma_P$ and $\Gamma/\Gamma_\text{obs}$ above and to the left of that line could be recovered.}
    \label{fig:vortex_recovery}
\end{figure}

Finally, we must interpret these results in terms of our ability to recover vortices using our matched filter approach. In particular, we need to know the best shape for the matched filter: too narrow a filter might miss wider vortex profiles, while a wide filter could average out narrow profiles. To that end, we generated 500,000 synthetic time-series (200 distinct simulations for each of 2500 combinations of model parameters discussed below). These time-series had the same sampling as the APSS time-series and white noise with a wide range of variances $\sigma_P$. Into these time-series, we injected vortex signals with known depths and widths. Then, we applied a matched filter with width $\Gamma$ to see the range of values we retrieved for the convolution of the filter against the synthetic time-series, $\left( F \ast P \right)$. Figure \ref{fig:vortex_recovery} illustrates the range of such values and indicates that, for a very wide range of vortices, noise levels, and matched filter widths, we can successfully recover the vortices. Indeed, Figure \ref{fig:vortex_recovery} shows we could even recover very subtle vortices, given the right filter width. For example, for $\Delta P_{\rm obs}/\sigma_P$ (i.e., a vortex that barely rises above the noise), a wide range of filter widths $\Gamma$ returns $\left( F \ast P \right) \geq 5 $. The noise model discussed here does not include the non-white (red) noise that pervades the real APSS data, meaning the results are somewhat optimistic. However, based on the results here (and on additional experimentation with the real APSS time-series), we chose $\left( F \ast P \right) \geq 5 $ as our vortex detection threshold. 

By design, our detection scheme will filter out some vortex signals. In particular, vortices with pressure signals very different from a Lorentzian will be missed. As an example, a Vatistas vortex that passes over the sensor in a non-linear trajectory would not generate a Lorentzian; however, such encounters seem to be unusual \citep{LORENZ20151}, so we do not consider them. So what about simple Lorentzians -- how many vortices of a given pressure drop is our approach likely to have missed? A simple way to address this question is to consider how often the pressure time-series were too noisy to have detected a vortex of a given $\Delta P_{\rm obs}$. Figure \ref{fig:vortex_recovery} suggests that, for most of the vortices we consider, a threshold $F \ast P \ge 5$ requires $\log_{10} \left( \Delta P_{\rm obs}/\sigma_P \right) \ge -0.5$. For the vortex with the smallest $\Delta P_{\rm obs} = 0.1\,{\rm Pa}$, this requirement translates to $\sigma_P \le 0.3\,{\rm Pa}$. On sols with scatter larger than that threshold, we could not (in principle) have detected such vortices. Of the sols we analyzed, only about 18\% had such large scatter, meaning our approach likely missed few such vortices. For more typical vortices (the median $\Delta P_{\rm obs} = 1.1\,{\rm Pa}$), none of our roughly 400 sols had sufficiently high scatter that we would have failed to detect the vortex, suggesting a miss rate of much less than 1 in 400 for vortex signals matching our detection criteria.

\section{Inferring Encounter Geometries from the Pressure and Velocity Profiles}
\label{sec:Inferring Encounter Geometries from the Pressure and Velocity Profiles}
We assume the vortices correspond to Vatistas vortices \citep{1991ExFl...11...73V}, giving pressure $\Delta P$ and tangential wind velocity profiles $V$:
\begin{equation}
    \Delta P(r) = -\frac{\Delta P_{\rm act}}{1 + \left( \frac{r}{D_{\rm act}/2} \right)^2}\label{eqn:radial_lorentzian_profile_appendix}
\end{equation}
and
\begin{equation}
    V(r) = \frac{ V_{\rm act} \left( 2 \frac{r}{D_{\rm act}/2} \right)}{1 + \left( \frac{r}{D_{\rm act}/2} \right)^2},\label{eqn:radial_wind_profile}
\end{equation}
where $r$ is the radial distance from the vortex center, $P_{\rm act}$ the central pressure excursion, and $V_{\rm act}$ the peak velocity at a distance $D_{\rm act}$ from the vortex center. Equation \ref{eqn:radial_distance} shows the evolution of the radial distance with time during an encounter. At closest approach $r = b$, and we measure $\Delta P = \Delta P_{\rm obs}$ and 
\begin{equation}
    V = \frac{ V_{\rm act} \left( 2\frac{b}{D_{\rm act}/2} \right)}{1 + \left( \frac{b}{D_{\rm act}/2} \right)^2} = V_{\rm obs}.\label{eqn:Vobs}
\end{equation} 

Cyclostrophic balance relates the pressure and velocity \citep{2020Icar..33813523J} as:
\begin{equation}
    V_{\rm act} = \left( \Delta P_{\rm act}/\rho \right)^{1/2},\label{eqn:cyclostrophic_balance}
\end{equation}
where $\rho$ is the atmospheric density. Combining Equations \ref{eqn:radial_lorentzian_profile_appendix}, \ref{eqn:Vobs}, and \ref{eqn:cyclostrophic_balance} gives
\begin{equation}
    \Delta P_{\rm act} = \left( 1 - \frac{\rho V_{\rm obs}^2}{4 \Delta P_{\rm obs}} \right)^{-1} \Delta P_{\rm obs},\label{eqn:solution_for_Pact}
\end{equation}
and
\begin{equation}
    V_{\rm act} = \left( 1 - \frac{\rho V_{\rm obs}^2}{4 \Delta P_{\rm obs}} \right)^{-1/2} \left( \Delta P_{\rm obs}/\rho \right)^{1/2}.\label{eqn:solution_for_Vact}
\end{equation}

We can then solve for $D_{\rm act}$:
\begin{equation}
    D_{\rm act} = \sqrt{ D_{\rm obs}^2 - \left( 2 b\right)^2 }.\label{eqn:actual_diameter}
\end{equation}

To determine the uncertainties on the actual parameters, we propagate uncertainties from the model-fit observed parameters $\sigma_{\Delta P_{\rm obs}}$, $\sigma_{D_{\rm obs}}$, $b$, and $\sigma_{V_{\rm obs}}$:
\begin{equation}
    \sigma_{\rm \Delta P_{\rm act}} = \sqrt{ \left( \frac{\partial \Delta P_{\rm act}}{\partial \Delta P_{\rm obs}} \sigma_{\Delta P_{\rm obs}} \right)^2 + \left( \frac{\partial \Delta P_{\rm act}}{\partial V_{\rm obs}} \sigma_{V_{\rm obs}} \right)^2 } = \left( \frac{\Delta P_{\rm act}}{\Delta P_{\rm obs}} \right) \sqrt{ \frac{1}{4} \rho^2 V_{\rm obs}^2 \sigma_{\Delta P_{\rm obs}}^2 + \left[ 1 + \left( \frac{\Delta P_{\rm act}}{\Delta P_{\rm obs}} \right) \left( \frac{\rho V_{\rm obs}^2}{4 \Delta P_{\rm obs}^2} \right) \right]^2 \sigma_{V_{\rm obs}}^2 }
\end{equation}
\begin{equation}
    \sigma_{\rm V_{\rm act}} = \left( \frac{\partial V_{\rm act}}{\partial \Delta P_{\rm act}} \right) \sigma_{\Delta P_{\rm act}} = \left( \frac{V_{\rm act}}{2\Delta P_{\rm act}} \right) \sigma_{\Delta P_{\rm act}}
\end{equation}
\begin{equation}
    \sigma_{D_{\rm act}} = \sqrt{ \left( \frac{\partial D_{\rm act}}{\partial D_{\rm obs}} \sigma_{D_{\rm obs}} \right)^2 + \left( \frac{\partial D_{\rm act}}{\partial b} \sigma_{b}\right)^2 } = \sqrt{ \left( \frac{D_{\rm obs}}{D_{\rm act}} \sigma_{D_{\rm obs}}\right)^2 + \left( \frac{2b}{D_{\rm act}} \sigma_{b} \right)^2 }
\end{equation}

\bibliography{sample63}{}
\bibliographystyle{aasjournal}

\end{document}